\documentclass[twocolumn]{aastex63}

\usepackage{natbib}
\usepackage{color}
\usepackage{mathtools}
\usepackage{hyperref} 

\def    \apjl  		{\rm {ApJL}}

\def    \apjl  		{\rm {ApJL}}

\def	\cm		{\,{\rm {cm}}}
\def	\K		{\,{\rm K}}
\def	\g		{\,{\rm {g}}}
\def	\mum	{\,{\mu \rm{m}}}

\def \bea {\begin{eqnarray}}
\def \ena {\end{eqnarray}}

\def	\B	{{\rm B}}

\def	\C	{{\rm C}}

\def	\cm	{\,{\rm cm}}
\def	\d	{{\rm d}}

\def	\D	{{\rm D}}

\def	\erg	{\,{\rm erg}}

\def	\g	{\,{\rm g}}
\def	\gas	{\,{\rm gas}}
\def	\G	{{\rm G}}

\def	\H	{{\rm H}}

\def	\IR	{{\rm IR}}

\def	\N	{{\rm N}}

\def	\s	{\,{\rm s}}

\def	\V	{{\rm V}}

\def	\rad	{{\rm rad}}

\def    \gas     	{{\rm gas}}

\begin{document}
\shorttitle{Dust evolution with redshift}
\shortauthors{Thiem Hoang}
\title{Variation of dust properties with cosmic time implied by radiative torque disruption}

\author{Thiem Hoang}
\affiliation{Korea Astronomy and Space Science Institute, Daejeon 34055, Republic of Korea, \href{mailto:thiemhoang@kasi.re.kr}{thiemhoang@kasi.re.kr}}
\affiliation{Korea University of Science and Technology, 217 Gajeong-ro, Yuseong-gu, Daejeon, 34113, Republic of Korea}

\begin{abstract}
Dust properties within a galaxy are known to change from the diffuse medium to dense clouds due to increased local gas density. However, the question of whether dust properties change with redshift remains elusive. In this paper, using the fact that the mean radiation intensity of the interstellar medium (ISM) of star-forming galaxies increases with redshift, we show that dust properties should change due to increasing efficiency of rotational disruption by radiative torques, an effect named RAdiative Torque Disruption (RATD). We first show that, due to RATD, the size distribution of interstellar dust varies with redshift, such as dust grains become smaller at higher $z$. We model the extinction curves and find that the curve becomes steeper with increasing redshift. The ratio of total-to-selective extinction, $R_{V}$, decreases with redshift and achieves low values of $R_{V}\sim 1.5-2.5$ for grains having a composite structure. We also find that dust properties change with the local gas density due to RATD, but the change is dominated by the radiation field for the diffuse ISM. The low values of $R_{V}$ implied by RATD of interstellar dust could reproduce anomalous dust extinction observed toward type Ia supernovae (SNe Ia) and Small Magellanic Cloud (SMC)-like extinction curves with a steep far-UV rise toward high-z galaxies. Fluctuations in $R_{V}$ due to interstellar turbulence and varying radiation intensity may resolve the tension in measurements of the Hubble constant using SNe Ia. We finally discuss the implications of evolving dust properties for high-z astrophysics. 
\end{abstract}

\section{Introduction}\label{sec:intro}
Interstellar dust is an essential component of the interstellar medium (ISM) and plays critical roles in astrophysics, including gas heating, star and planet formation, and grain-surface chemistry (see \citealt{1990ARA&A..28...37M} and \citealt{2003ARA&A..41..241D} for reviews). Dust extinction and emission are key for extragalactic astrophysics, including measurements of star formation efficiency and understanding cosmic evolution (\citealt{2001PASP..113.1449C}; see \citealt{2020arXiv200103181S} for a review). Accurate measurements of star formation rate (SFR) depend crucially on dust properties, including grain size distribution, shape, and composition. Among the different dust properties, the grain size distribution is the most important parameter that determines emission, extinction, and polarization of dust.

Following the popular paradigm of dust evolution, {\it stardust} grains form in the dense outflows of evolved stars and in the dense ejecta of core-collapse supernovae, which dominantly produce large grains (\citealt{2003ApJ...598..785N}). Such large grains are then fragmented into small grains by shattering in shocks when being released into the diffuse ISM (\citealt{1994ApJ...433..797J}). Subsequently, interstellar dust is reprocessed in the ISM by growth (gas accretion and coagulation) and destruction processes, including sublimation, sputtering and shattering in shocks. Thus, the size distribution of interstellar dust is determined by the balance between growth and destruction processes in the ISM. On the other hand, dust properties are expected to change from the diffuse ISM (with gas density less than $100\cm^{-3}$) to dense molecular clouds (density above $100\cm^{-3}$) due to gas accretion and grain coagulation (\citealt{2016ApJ...831..147Z}). This process is mostly determined by gas density that controls gas-grain and grain-grain collision rates. The above dust evolution cycle inevitably changes the internal structure of grains, from a presumably compact structure for stardust to a composite structure for interstellar dust (\citealt{1990ARA&A..28...37M}; \citealt{Draine:2020ua}). Therefore, dust properties in the ISM of a galaxy are expected to change. The question now is, assuming the same gas density of the ISM, whether and how dust properties change with cosmic time (i.e., redshift). 

Recently, \cite{Hoang:2019da} introduced a new physical mechanism of dust destruction that changes the size distribution of interstellar grains due to radiative torques (RATs), which was termed Radiative Torque Disruption (RATD). The basic idea of the RATD mechanism is that, dust grains of irregular shape, when irradiated by anisotropic radiation field, experiences RATs (\citealt{1976Ap&SS..43..291D}; \citealt{1996ApJ...470..551D}; \citealt{2007MNRAS.378..910L}) and can be spun-up to extremely fast rotation (\citealt{1996ApJ...470..551D}; \citealt{2009ApJ...695.1457H}). Resulting centrifugal stress can exceed the maximum tensile strength of grain material, resulting in the disruption of the grain into fragments. Previously, \cite{2016ApJ...818..133S} noticed that interplanetary grains of fluffy structure could be disrupted by RATs induced by solar radiation. Since rotational disruption acts to break loose bonds between the grain constituents, unlike breaking strong chemical bonds between atoms in thermal sublimation, RATD can work with the average interstellar radiation field (see \citealt{2020Galax...8...52H} for a review). The RATD mechanism introduces a new environment parameter for dust evolution, namely local radiation intensity, and is found to be the most efficient mechanism that constrains the upper limit of the size distribution (\citealt{2019ApJ...876...13H}). Simulations of grain evolution with cosmic time in \cite{2020MNRAS.494.1058H} show that RATD is indeed the key factor determining the upper cutoff of the size distribution. The efficiency of RATD depends on the local conditions, including gas density and radiation field, grain properties (internal structure, size, and shape) \citep{2019ApJ...876...13H}, and grain alignment (\citealt{Lazarian:2020ta}). Therefore, if the density and radiation strength vary with redshift, the local environment changes, dust properties should change accordingly.

It is well-known that star formation rate (SFR) increases with redshift (see \citealt{Bethermin:2015hn} and references therein). Since the mean intensity of the ISM of galaxies is governed by star formation activity, it is expected to increase with $z$. Numerous observations reveal the increase of dust temperature with redshift (\citealt{1999ApJ...517L..19O}; \citealt{2012ApJ...760....6M}; \citealt{2017ApJ...847...21F}; \citealt{Bethermin:2015hn}; \citealt{2017MNRAS.471.5018F}; \citealt{2017MNRAS.472.4587H}; \citealt{2020arXiv200409528S}). Theoretical studies (\citealt{delPLagos:2012hh}; \citealt{2017MNRAS.467.1231C}; \citealt{2018ApJ...854...36I}) and cosmological simulations (\citealt{2018MNRAS.474.1718N}) imply the increase of the mean radiation intensity with redshift. \cite{2010MNRAS.409...75H} also found the increase of $T_{d}$ with redshift using Herschel. Other studies of star-forming galaxies also report the increase of the mean radiation intensity with redshift (\citealt{2012ApJ...760....6M}; \citealt{2020ApJ...889...80L}). Thus, the interstellar radiation field is stronger in higher-z galaxies than in the Galaxy. As a result, interstellar grains experience stronger RATs, and rotational disruption is more efficient, resulting in smaller grains with increasing $z$. We will quantify this effect using the established relationship of the mean radiation intensity with $z$ and explore its observational consequences.

The structure of the present paper is as follows. In Section \ref{sec:radfield}, we describe the interstellar radiation field of galaxies at different redshift inferred from observations that will be used for our study. In Section \ref{sec:RATD}, we briefly review the basic features of Radiative Torques (RATs) and rotational disruption by radiative torques (RATD) mechanism, and derive the critical grain size for rotational disruption. In Section \ref{sec:ext}, we present numerical results for the disruption size and calculate resulting extinction curves using the size distribution constrained by RATD for different redshift and local gas density. In Section \ref{sec:disc}, we discuss the implications of our results for understanding anomalous dust properties observed toward type Ia supernovae (SNe Ia) and high-z astrophysics. A summary of our main findings is given in Section \ref{sec:summ}.

\section{Interstellar radiation field of galaxies}\label{sec:radfield}

Let $u_{\lambda}$ be the spectral energy density of radiation field at wavelength $\lambda$. The energy density of the radiation field is then $u_{\rad}=\int_{0}^{\infty} u_{\lambda}d\lambda$. To describe the strength of a radiation field, let define $U=u_{\rm rad}/u_{\rm ISRF}$ with 
$u_{\rm ISRF}=8.64\times 10^{-13}\erg\cm^{-3}$ being the energy density of the average interstellar radiation field (ISRF) in the solar neighborhood as given by \cite{1983A&A...128..212M}. Thus, the typical value for the ISRF in the solar neighborhood is $U=1$.

For star-forming galaxies, \cite{Bethermin:2015hn} derived the best-fit for the increase of the mean radiation intensity with redshift from $z=0-4$ as
\bea
U = U_{0}(1+z)^{\alpha_{z}}\label{eq:U_z}
\ena
where $U_{0}=3\pm 1.1$ is the mean intensity at $z=0$ and $\alpha_{z}=1.8\pm 0.4$ is the power-law index (\citealt{2017A&A...603A..93M}; \citealt{2018A&A...609A..30S}).



Recent analysis from \cite{Bouwens:2020tf} obtained a best-fit to observational data for the dust temperature as
\bea
T_{d} = (34.6 \pm 0.3) + (3.94 \pm 0.26)(z - 2)\K,\label{eq:Td_z}
\ena
for $z\sim 0-10$, which corresponds to the increase in the mean radiation intensity of $ U =(T_{d}/18.2\K)^{5.57}$ \citep{2018A&A...609A..30S}. 

In the following, we use Equation (\ref{eq:U_z}) for $z<4$ and Equation (\ref{eq:Td_z}) for $z>4$.

\section{Review of Rotational Disruption by Radiative Torques}\label{sec:RATD}

\subsection{Radiative torques of irregular grains}
Dust grains of irregular shape irradiated by an anisotropic radiation experience Radiative Torque (RAT). The magnitude of RAT is defined as
\bea
{\Gamma}_{\lambda}=\pi a^{2}
\gamma u_{\lambda} \left(\frac{\lambda}{2\pi}\right){Q}_{\Gamma},\label{eq:GammaRAT}
\ena
where $\gamma$ is the anisotropy degree of the radiation field, ${Q}_{\Gamma}$ is the RAT efficiency, and $a$ is the effective size of the grain which is defined as the radius of the sphere with the same volume as the irregular grain (\citealt{1996ApJ...470..551D}; \citealt{2007MNRAS.378..910L}).

The magnitude of RAT efficiency, $Q_{\Gamma}$, can be approximated by a power-law (\citealt{Hoang:2008gb}):
\bea
Q_{\Gamma}\approx 0.4\left(\frac{{\lambda}}{1.8a}\right)^{\eta},\label{eq:QAMO}
\ena
where $\eta=0$ for $\lambda \lesssim 1.8a$  and $\eta=-3$ for $\lambda > 1.8a$. 

Numerical calculations of RATs for several shapes of different optical constants in \cite{2007MNRAS.378..910L} find the slight difference in RATs among the realization. An extensive study for a large number of irregular shapes by \cite{2019ApJ...878...96H} shows little difference in RATs for silicate, carbonaceous, and iron compositions. Moreover, the analytical formula (Equation \ref{eq:QAMO}) is also in a good agreement with their numerical calculations. Therefore, one can use Equation (\ref{eq:QAMO}) for the different grain compositions and grain shapes, and the difference is an order of unity

Let $\bar{\lambda}=\int_{0}^{\infty} \lambda u_{\lambda}d\lambda/u_{\rm rad}$ be the mean wavelength of the radiation field. For the spectrum of the ISRF in our galaxy, $\bar{\lambda}=1.2\mum$ (\citealt{Hoang:2020vg}). We assume that the spectrum of the ISRF of galaxies is similar to the Galaxy, so that the value of $\bar{\lambda}$ remains the same with redshift. In reality, $\bar{\lambda}$ is expected to be smaller for higher $z$ due to bluer radiation emitted by more massive stars.

The average radiative torque efficiency over the spectrum is defined as
\bea
\overline{Q}_{\Gamma} = \frac{\int_{0}^{\infty} \lambda Q_{\Gamma}u_{\lambda} d\lambda}{\int_{0}^{\infty} \lambda u_{\lambda} d\lambda},
\ena
where the integrals are taken over the entire radiation spectrum.

For interstellar grains with $a\lesssim \overline{\lambda}/1.8$, $\overline{Q}_{\Gamma}$ can be approximated to (\citealt{2014MNRAS.438..680H})
\bea
\overline{Q}_{\Gamma}\simeq 2\left(\frac{\overline{\lambda}}{a}\right)^{-2.7}\simeq 2.6\times 10^{-2}\left(\frac{\overline{\lambda}}{0.5\mum}\right)^{-2.7}a_{-5}^{2.7},
\ena
where $a_{-5}=a/(10^{-5}\cm)$, and $\overline{Q_{\Gamma}}\sim 0.4$ for $a> \overline{\lambda}/1.8$. \cite{Hoang:2020vg} used rigorous mathematical derivations for the mean RAT and find that the above scaling is a good fit to numerical calculations for the ISRF.  

Therefore, the average RAT can be given by
\bea
\Gamma_{\rm RAT}&=&\pi a^{2}
\gamma u_{\rad} \left(\frac{\overline{\lambda}}{2\pi}\right)\overline{Q}_{\Gamma}\nonumber\\
&\simeq & 5.8\times 10^{-29}a_{-5}^{4.7}\gamma U\overline{\lambda}_{0.5}^{-1.7}\erg,~~~
\ena
for $a\lesssim \overline{\lambda}/1.8$, and
\bea
\Gamma_{\rm RAT}\simeq & 8.6\times 10^{-28}a_{-5}^{2}\gamma U\overline{\lambda}_{0.5}\erg,~~~
\ena
for $a> \bar{\lambda}/1.8$, where the mean wavelength is normalized over the optical wavelength, $\overline{\lambda}_{0.5}=\overline{\lambda}/0.5\mum$

The well-known damping process for a rotating grain is sticking collision with gas species (atoms and molecules), followed by their thermal evaporation. Thus, for a gas with He of $10\%$ abundance, the characteristic damping time is
\bea
\tau_{\gas}&=&\frac{3}{4\sqrt{\pi}}\frac{I}{1.2n_{\rm H}m_{\rm H}
v_{\rm th}a^{4}}\nonumber\\
&\simeq& 8.74\times 10^{4}a_{-5}\hat{\rho}\left(\frac{30\cm^{-3}}{n_{\H}}\right)\left(\frac{100\K}{T_{\gas}}\right)^{1/2}~{\rm yr},~~
\ena
where $I=8\pi \rho a^{5}/15$ is the grain inertia moment of spherical grain of effective radius $a$, $\hat{\rho}=\rho/(3\g\cm^{-3})$ with $\rho$ being the dust mass density, $v_{\rm th}=\left(2k_{\B}T_{\rm gas}/m_{\rm H}\right)^{1/2}$ is the thermal velocity of a gas atom of mass $m_{\rm H}$ in a plasma with temperature $T_{\gas}$ and density $n_{\H}$ (\citealt{1996ApJ...470..551D}; \citealt{2009ApJ...695.1457H}). The gas damping time is estimated for spherical grains, and we disregard the factor of unity due to grain shape. 

Infrared (IR) photons emitted by the grain carry away part of the grain's angular momentum, resulting in the damping of the grain rotation. For strong radiation fields or not very small sizes, grains can achieve equilibrium temperature, such that the IR damping coefficient (see \citealt{1998ApJ...508..157D}) can be calculated as
\bea
F_{\rm IR}\simeq \left(\frac{0.4U^{2/3}}{a_{-5}}\right)
\left(\frac{30 \cm^{-3}}{n_{\H}}\right)\left(\frac{100 \K}{T_{\gas}}\right)^{1/2}.\label{eq:FIR}
\ena 

Other rotational damping processes include plasma drag, ion collisions, and electric dipole emission. These processes are mostly important for polycyclic aromatic hydrocarbons (PAHs) and very small grains of radius $a<0.01\mum$ (\citealt{1998ApJ...508..157D}; \citealt{Hoang:2010jy}; \citealt{2011ApJ...741...87H}). Thus, the total rotational damping rate by gas collisions and IR emission can be written as
\bea
\tau_{\rm damp}^{-1}=\tau_{\gas}^{-1}(1+ F_{\rm IR}).\label{eq:taudamp}
\ena

For strong radiation fields of $U\gg 1$ and not very dense gas, one has $F_{\rm IR}\gg 1$. Therefore, $\tau_{\rm damp}\sim \tau_{\gas}/F_{\IR}\sim a_{-5}^{2}U^{2/3}$, which does not depend on the gas properties. In this case, the only damping process is caused by IR emission.

For radiation sources with stable luminosity considered in this paper, radiative torque, $\Gamma_{\rm RAT}$, is constant, and the grain velocity is steadily increased over time. The equilibrium rotation can be achieved at (see \citealt{2007MNRAS.378..910L}; \citealt{2009ApJ...695.1457H}; \citealt{2014MNRAS.438..680H}):
\bea
\omega_{\rm RAT}=\frac{\Gamma_{\rm RAT}\tau_{\rm damp}}{I}.\label{eq:omega_RAT0}
\ena

The rotation rate by RATs is given by
\bea
\omega_{\rm RAT}
&=& \frac{5a^{0.7}\gamma u_{\rm rad}\bar{\lambda}^{-1.7}}{8n_{\rm H}\sqrt{2\pi m_{\rm H}kT_{\rm gas}}}\left(\frac{1}{1+F_{\rm IR}}\right)\nonumber\\
&\simeq &9.22\times 10^{7}a_{-5}^{0.7}\bar{\lambda}_{0.5}^{-1.7}\nonumber\\
&\times&\left(\frac{\gamma_{-1}U}{n_{1}T_{2}^{1/2}}\right)\left(\frac{1}{1+F_{\rm IR}}\right)\rad\s^{-1},~~~\label{eq:omega_RAT}
\ena
for grains with $a\lesssim \bar{\lambda}/1.8$, and
\bea
\omega_{\rm RAT}
&=&\frac{a^{-2}\gamma u_{\rm rad}\bar{\lambda}}{4n_{\rm H}\sqrt{2\pi m_{\rm H}kT_{\rm gas}}}\left(\frac{1}{1+F_{\rm IR}}\right)\nonumber\\
&\simeq &1.42\times 10^{9}a_{-5}^{-2}\bar{\lambda}_{0.5}\nonumber\\
&&\times \left(\frac{\gamma_{-1} U}{n_{1}T_{2}^{1/2}}\right)\left(\frac{1}{1+F_{\rm IR}}\right)\rad\s^{-1},\label{eq:omega_RAT2}
\ena
for grains with $a> \overline{\lambda}/1.8$. 

Above, $n_{1}=n_{\H}/(10\cm^{-3})$, $T_{2}=T_{\gas}/100\K$, $\gamma_{-1}=\gamma/0.1$ is the anisotropy of radiation field relative to the typical anisotropy of the diffuse interstellar radiation field. The radiation anisotropy degree also varies with the location, between $\gamma\sim 0.1$ for the diffuse ISM (\citealt{1997ApJ...480..633D}) to $\gamma\sim 0.7$ for molecular clouds (\citealt{2007ApJ...663.1055B}), and $\gamma=1$ for grains close to a star. For our study here for the diffuse ISM, we adopt the typical value of $\gamma=0.1$. The mean wavelength is fixed to $\bar{\lambda}=1.2\mum$, although it may decrease with redshift due to harder radiation field.

\subsection{Maximum grain size constrained by Radiative Torque Disruption}\label{sec:disr}
A spherical dust grain of radius $a$ rotating at velocity $\omega$ develops an average tensile stress due to centrifugal force which scales as (see \citealt{Hoang:2019da})
\bea
S=\frac{\rho a^{2} \omega^{2}}{4}.\label{eq:Stress}
\ena 

When the rotation rate is sufficiently high such as the tensile stress exceeds the maximum limit, namely tensile strength $S_{\rm max}$, the grain is disrupted. The critical rotational velocity is given by $S=S_{\rm max}$:
\bea
\omega_{\rm disr}&=&\frac{2}{a}\left(\frac{S_{\max}}{\rho} \right)^{1/2}\nonumber\\
&\simeq& \frac{3.6\times 10^{8}}{a_{-5}}S_{\max,7}^{1/2}\hat{\rho}^{-1/2}~\rad\s^{-1},\label{eq:omega_cri}
\ena
where $S_{\max,7}=S_{\max}/10^{7} \erg \cm^{-3}$ (\citealt{Hoang:2019da}).
 
The tensile strength of interstellar dust depends on grain structure (compact vs. composite vs core-mantle), which is uncertain (\citealt{1990ARA&A..28...37M}). Compact grains should have large tensile strength of $S_{\rm max}\gtrsim 10^{9}\erg\cm^{-3}$, whereas composite/fluffy grains have much lower tensile strength \citep{2019ApJ...876...13H}. Large grains (radius $a>0.1\mum$) are expected to have composite structure as a result of coagulation process in molecular clouds or in the ISM. Numerical simulations for porous grain aggregates from \cite{2019ApJ...874..159T} find that the tensile strength decreases with increasing the monomer radius and can be fitted with an analytical formula (see \citealt{2020MNRAS.496.1667K} for more details)
\bea
S_{\max} &\simeq& 9.51\times 10^{4} \left(\frac{\gamma_{\rm sf}}{0.1 J m^{-2}}\right) \nonumber\\
&\times&\left(\frac{r_{0}}{0.1\mum}\right)^{-1}\left(\frac{\phi}{0.1}\right)^{1.8} \erg\cm^{-3},\label{eq:Smax}
\ena
where $\gamma_{\rm sf}$ is the surface energy per unit area of the material, $r_{0}$ is the monomer radius, and $\phi$ is the volume filling factor of monomers. For large grains ($a>0.1\mum$) made of monomers of radius $r_{0}=0.1\mum$ and $\phi=0.1$, Equation (\ref{eq:Smax}) implies $S_{\max}\approx 10^{5}\erg\cm^{-3}$. 

Throughout this paper, we assume that large grains have composite structure, as expected from grain evolution model in the ISM (\citealt{1990ARA&A..28...37M}) and adopt $r_{0}=0.1\mum$, yielding the typical tensile strength of $S_{\rm max}=10^{5}\erg\cm^{-3}$. We also explore the possibilities that grains are made of smaller monomers ($r_{0}<0.1\mum$) or have core-mantle and compact structures with larger tensile strength of $S_{\rm max}=10^{6}-10^{8}\erg\cm^{-3}$.

For an arbitrary radiation field and $a\le \overline{\lambda}/1.8$, one obtains
\bea
a_{\rm disr}&\simeq&
\left(\frac{16n_{\H}\sqrt{2\pi m_{\H}kT_{\gas}}}{5\gamma u_{\rm rad}\bar{\lambda}^{-1.7}}\right)^{1/1.7}\left(\frac{S_{\max}}{\rho}\right)^{1/3.4}\nonumber\\
&\simeq0.22&\bar{\lambda}_
{0.5}S_{\max,7}^{1/3.4}(1+F_{\rm IR})^{1/1.7}\left(\frac{n_{1}T_{2}^{1/2}}{\gamma_{-1}U}\right)^{1/1.7}\mum,~~~~~\label{eq:adisr_U}
\ena
which depends only on the local gas density and temperature. The disruption size is the function of two parameters, gas density $n_H$ and the radiation strength.


Using the relationship between $U$ and redshift (Eq. \ref{eq:U_z}), one obtains
\bea
a_{\rm disr}&\simeq&0.22\bar{\lambda}_
{0.5}S_{\max,7}^{1/3.4} (1+F_{\rm IR})^{1/1.7}\nonumber\\
&&\times
\left(\frac{n_{1}T_{2}^{1/2}}{\gamma_{-1}}\right)^{1/1.7}(1+z)^{-\alpha_{z}/1.7}\mum,~~\label{eq:adisr_z}
\ena
where $\alpha_{z}\approx 1.8$. This equation implies the inversely decrease of the disruption size with redshift as $(1+z)^{-\alpha_{z}/1.7}\sim 1/(1+z)$. 

Due to the decrease of the rotation rate for $a>a_{\rm trans}$ (see Eq. \ref{eq:omega_RAT2}), there exist a maximum size, $a_{\rm disr,max}$, of grains that can still be disrupted by centrifugal stress (\citealt{2020ApJ...891...38H}). Setting $\omega_{\rm disr}=\omega_{\rm RAT}$ given by Equation (\ref{eq:omega_RAT2}) yields,
\bea
a_{\rm disr,max}&=&\frac{\gamma u_{\rm rad}\bar{\lambda}}{16n_{\rm H}\sqrt{2\pi m_{\rm H}kT_{\rm gas}}}\left(\frac{S_{\rm max}}{\rho}\right)^{-1/2}\nonumber\\
&\simeq& 
0.39\left(\frac{\gamma_{-1} U}{n_{1}T_{2}^{1/2}}\right)\bar{\lambda}_{0.5}\hat{\rho}^{1/2}S_{\max,7}^{-1/2}\nonumber\\
&\times&(1+F_{\rm IR})^{-1}\mum,\label{eq:adisr_up}
\ena 
which implies $a_{\rm disr,max}=3.9\mum$ for $S_{\rm max}=10^{5}\erg\cm^{-3}$ with $U=1$.

For strong radiation fields or low density such as $F_{IR}\propto U/n_{\H}>1$, IR damping dominates, then, the disruption size becomes independent of gas density,
\bea
a_{\rm disr}&\simeq& 2.4\left(\frac{\bar{\lambda}^{1.7}}{\gamma U^{1/3}}\right)^{1/2.7}\left(\frac{S_{\max}}{\rho}\right)^{1/5.4}\mum\nonumber\\
&\simeq&0.18\gamma_{-1}^{-1/2.7}U^{-1/8.1}\bar{\lambda}_
{0.5}^{1.7/2.7}(S_{\max,7}/\hat{\rho})^{1/5.4}\mum,~~~~~\label{eq:adisr_nogas1}
\ena
which corresponds to
\bea
a_{\rm disr}&\simeq&0.16\gamma_{-1}^{-1/2.7}(1+z)^{-\alpha_{z}/8.1}\bar{\lambda}_
{0.5}^{1.7/2.7}\nonumber\\
&&\times
(S_{\max,7}/\hat{\rho})^{1/5.4}\mum,~\label{eq:adisr_nogas2}
\ena
for $a_{\rm disr}\le \overline{\lambda}/1.8$.

In general, due to dependence of $F_{\rm IR}$ on the grain size $a$, one cannot obtain analytical $a_{\rm disr}$ as in Equation (\ref{eq:adisr_nogas1}). Thus, we first calculate numerically $\omega_{\rm RAT}$ using Equation (\ref{eq:omega_RAT}) and compare it with $\omega_{\rm disr}$ to find $a_{\rm disr}$ numerically, which will be referred to as numerical results. Note that the disruption size is assumed to be the same for silicate and carbonaceous grains. The results are shown in the next section.

\section{Grain Size Distribution and Extinction curves across cosmic time}\label{sec:ext}
\subsection{Grain size distribution}
The grain size distribution of dust is usually described by a power law,
\begin{align} \label{eq:dnda}
\frac{dn^{j}}{da} = C_{j} n_{\rm H} a^{\alpha},
\end{align} 
where $j$ denotes the grain composition (silicate and graphite), $C_{j}$ is the normalization constant, and $\alpha$ is the power slope. 

For the standard ISM in our galaxy, \cite{1977ApJ...217..425M} derived the slope $\alpha=-3.5$, $C_{\rm sil}=10^{-25.11}\cm^{2.5}$ for silicate grains, and $C_{\rm gra}=10^{-25.14}\cm^{2.5}$ for graphite grains. The size distribution has a lower cutoff of $a_{\rm min}=3.5$~\AA~ determined by thermal sublimation due to temperature fluctuations of very small grains (see e.g., \citealt{2007ApJ...663..866D}), and an upper cutoff of $a_{\rm max}=0.25\mum$ (\citealt{1977ApJ...217..425M}).
To account for the potential existence of large grains in the ISM, we assume $a_{\rm max, noRATD}=0.5\mum$ when RATD is not accounted for. 
In the presence of RATD, the maximum size $a_{\rm max}$ is determined by $\min(a_{\rm disr},a_{\max,\rm noRATD})$ because $a_{\rm disr,max}>0.5\mum$. Therefore, as $a_{\rm disr}$ changes with redshift due to RATD (see Equation \ref{eq:adisr_z}), the grain size distribution should change accordingly.

\begin{figure*}
\includegraphics[width=0.5\textwidth]{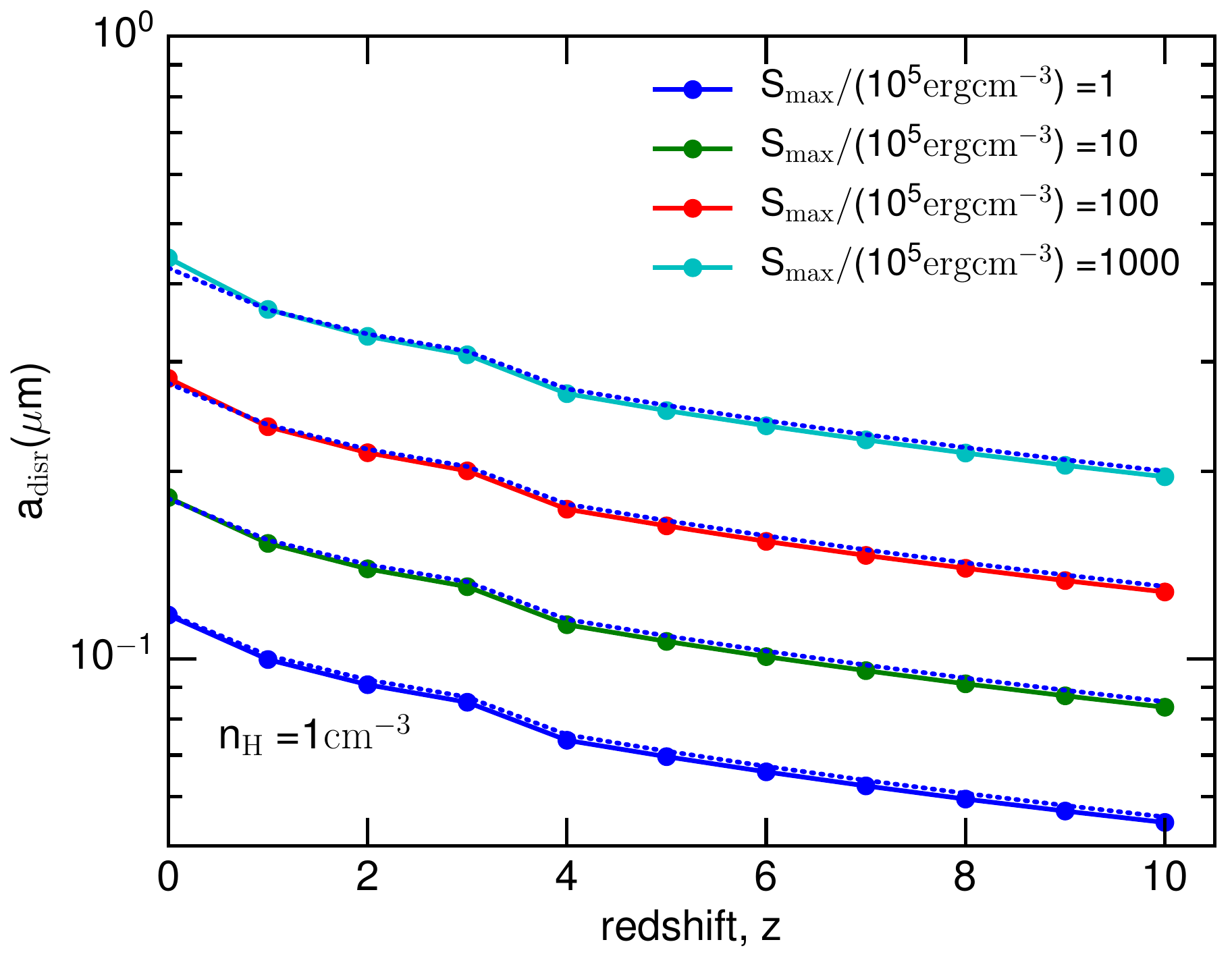}
\includegraphics[width=0.5\textwidth]{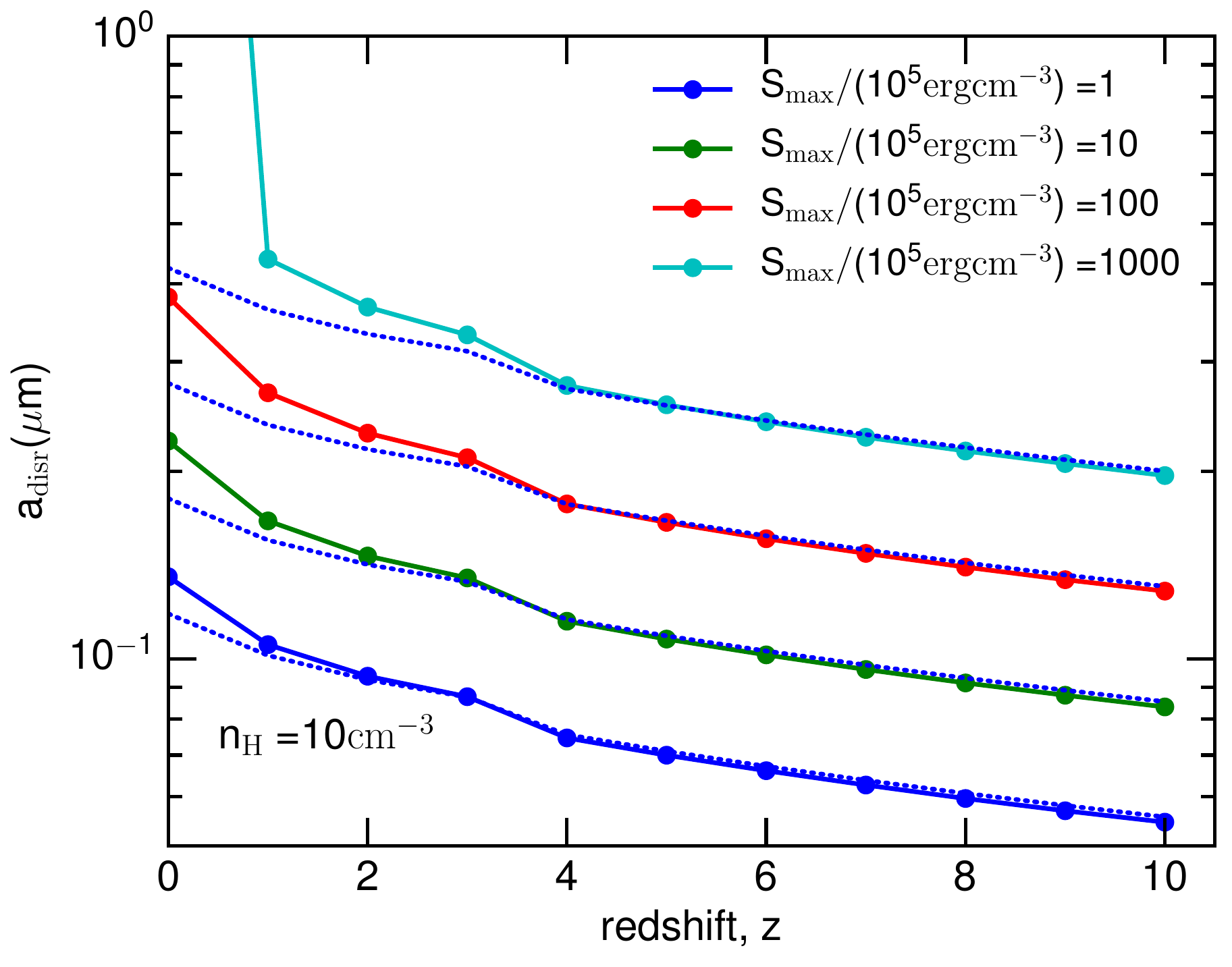}
\includegraphics[width=0.5\textwidth]{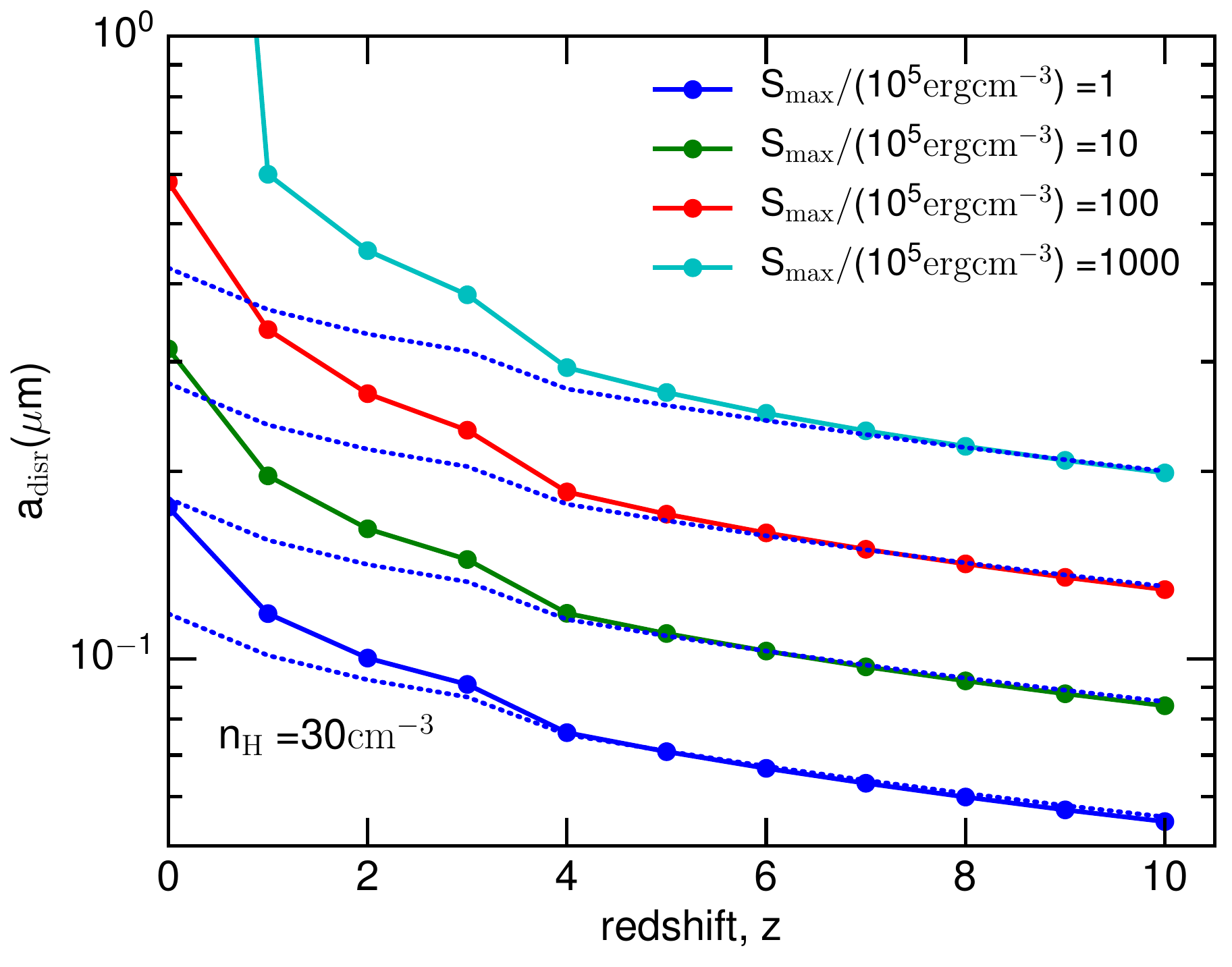}
\includegraphics[width=0.5\textwidth]{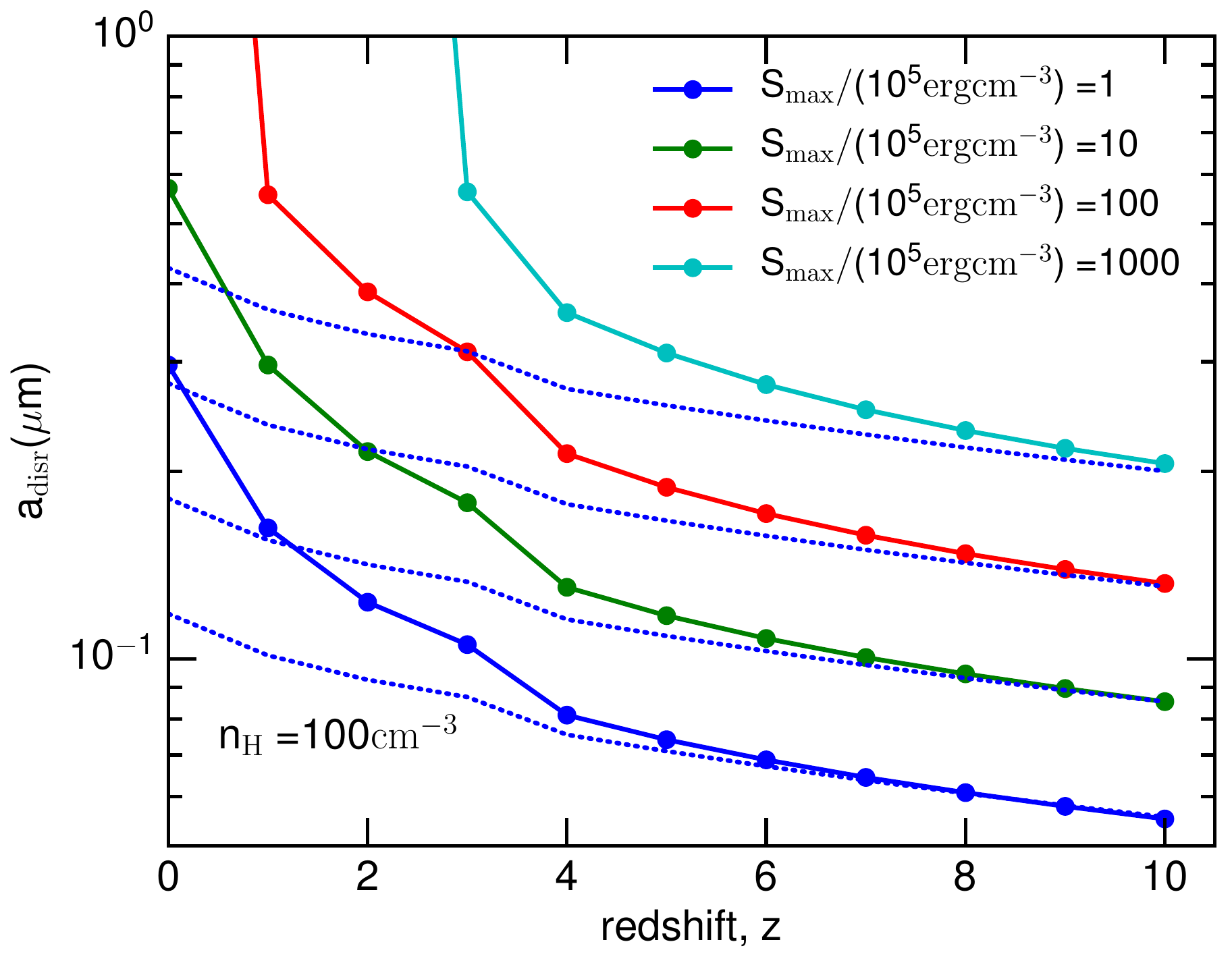}
\caption{Variation of disruption size with redshift for different tensile strength ($S_{\max}$) and gas density ($n_{\H}$). Analytical results in the absence of gas rotational damping from Equation (\ref{eq:adisr_nogas1}) are shown in dotted lines for comparison. The disruption size decreases with $z$, but increases with the tensile strength.}
\label{fig:adisr}
\end{figure*}

Figure \ref{fig:adisr} shows the variation of the disruption size with redshift for different density ($n_{\H}$) and tensile strength ($S_{\rm max}$) obtained from numerical calculations (solid lines) and analytical results where the gas damping is disregarded (dotted lines). We consider the maximum redshift of $z_{\max}=10$, corresponding to the age of $t_{\rm age}\sim 0.5$ Gyr. In general, analytical results obtained from Equation (\ref{eq:adisr_nogas1}) converge to numerical results for sufficiently high $z$ with large radiation intensity but are lower than the numerical results for high density and low $z$ due to the effect of gas damping. For a given density, $a_{\rm disr}$ decreases gradually with redshift. For the typical density of $n_{\H}=30\cm^{-3}$ and typical tensile strength of composite grains ($S_{\max}=10^{5}\erg\cm^{-3}$), the disruption size decreases from $a_{\rm disr}\sim 0.15\mum$ at $z=0$ to $0.1\mum$ at $z=2$, and $0.08\mum$ at $z=5$. The disruption size is lower for grains with a lower tensile strength, as implied by Equation (\ref{eq:adisr_U}). 

We also see that, for the low density cases (e.g., $n_{\H}=1, 10\cm^{-3}$), the disruption size changes slowly with $z$ as given by Equation (\ref{eq:adisr_nogas2}) due to dominance of IR damping. However, its change with $z$ is stronger for a higher density ($n_{\H}\sim 30, 100\cm^{-3}$) until the radiation field becomes sufficiently large for dominance of IR damping.

\subsection{Extinction curves and $R_{V}$}
The extinction of starlight by interstellar dust at wavelength $\lambda$ in the unit of magnitude per H is calculated as

\begin{align} \label{eq:Aext}
\frac{A_{\lambda}}{N_{\rm H}} = \sum_{j=\rm sil,gra} 1.086 \int_{a_{\rm min}}^{a_{\rm max}}  C_{\rm ext}^{j}(a)\left(\frac{1}{n_{\rm H}}\frac{dn^{j}}{da}\right)da,
\end{align}
where $N_{\rm H}=\int n_{\rm H}dz=n_{\rm H}L$ with $L$ the path length is the column density, $dn^{j}/da$ is the grain size distribution of dust component $j$ with the minimum size $a_{\rm min}$ and the maximum size $a_{\rm max}$ is taken to be $a_{\rm disr}$, $C_{\rm ext}^{j}$ is the cross-section of $j$ dust component which are calculated for oblate spheroidal grains of axial ratio of $2$ in \cite{2013ApJ...779..152H} assuming optical constant of astrosilicate and graphite \citep{1984ApJ...285...89D}.

Using the disruption size obtained in Figure \ref{fig:adisr}, we can calculate the wavelength-dependence extinction by dust being modified by RATD using Equation (\ref{eq:Aext}).

\begin{figure*}
\includegraphics[width=0.5\textwidth]{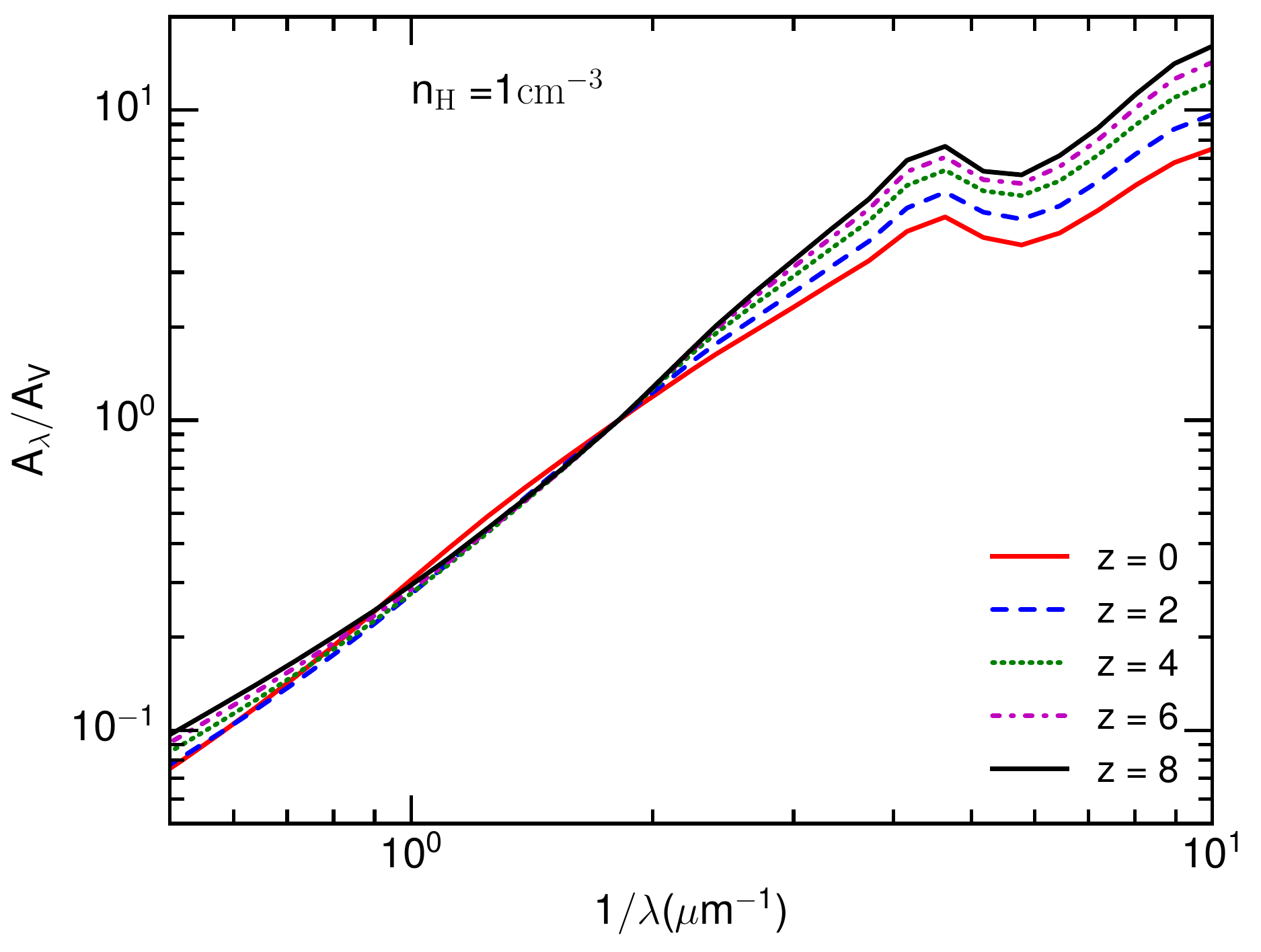}
\includegraphics[width=0.5\textwidth]{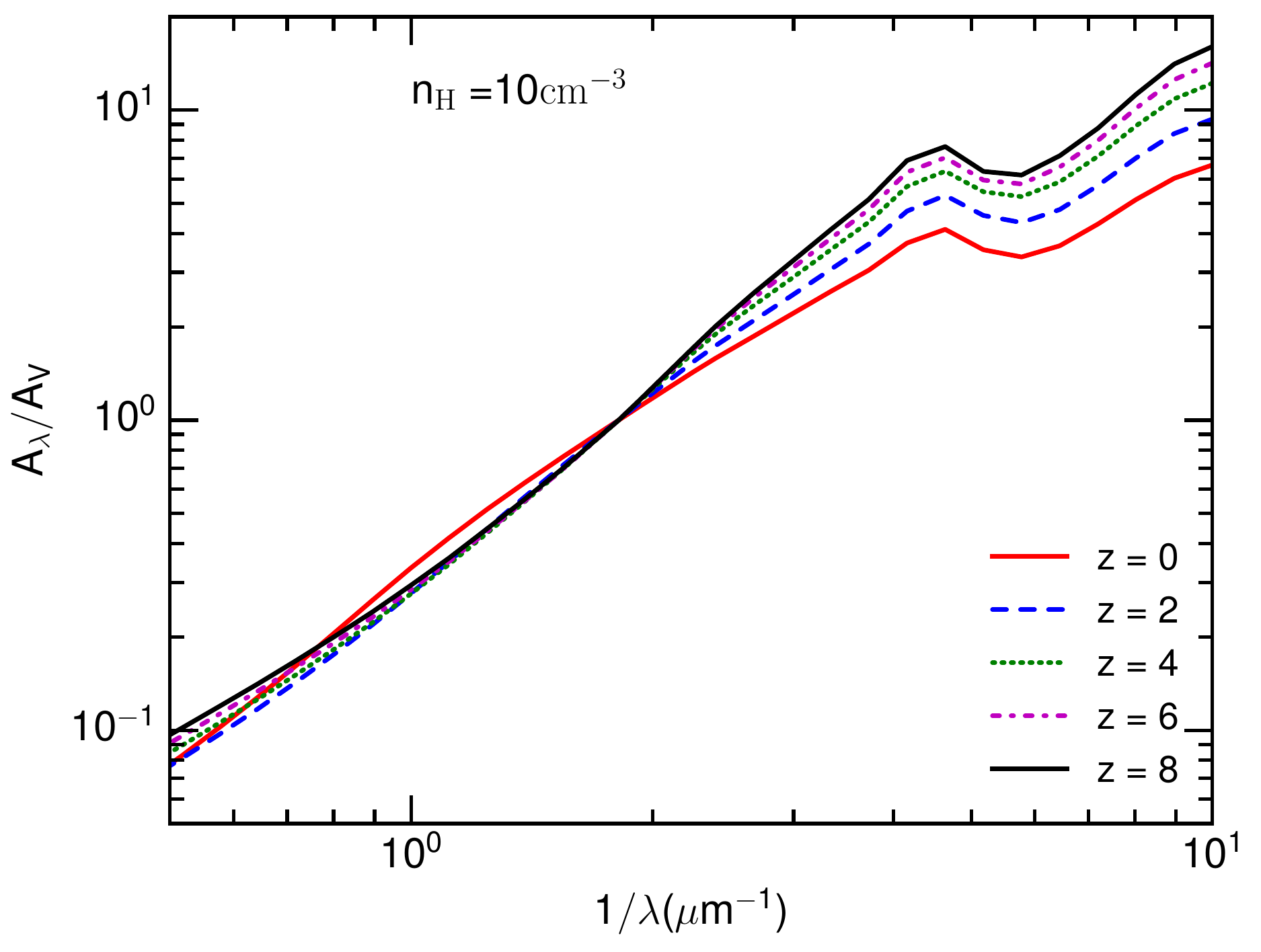}
\includegraphics[width=0.5\textwidth]{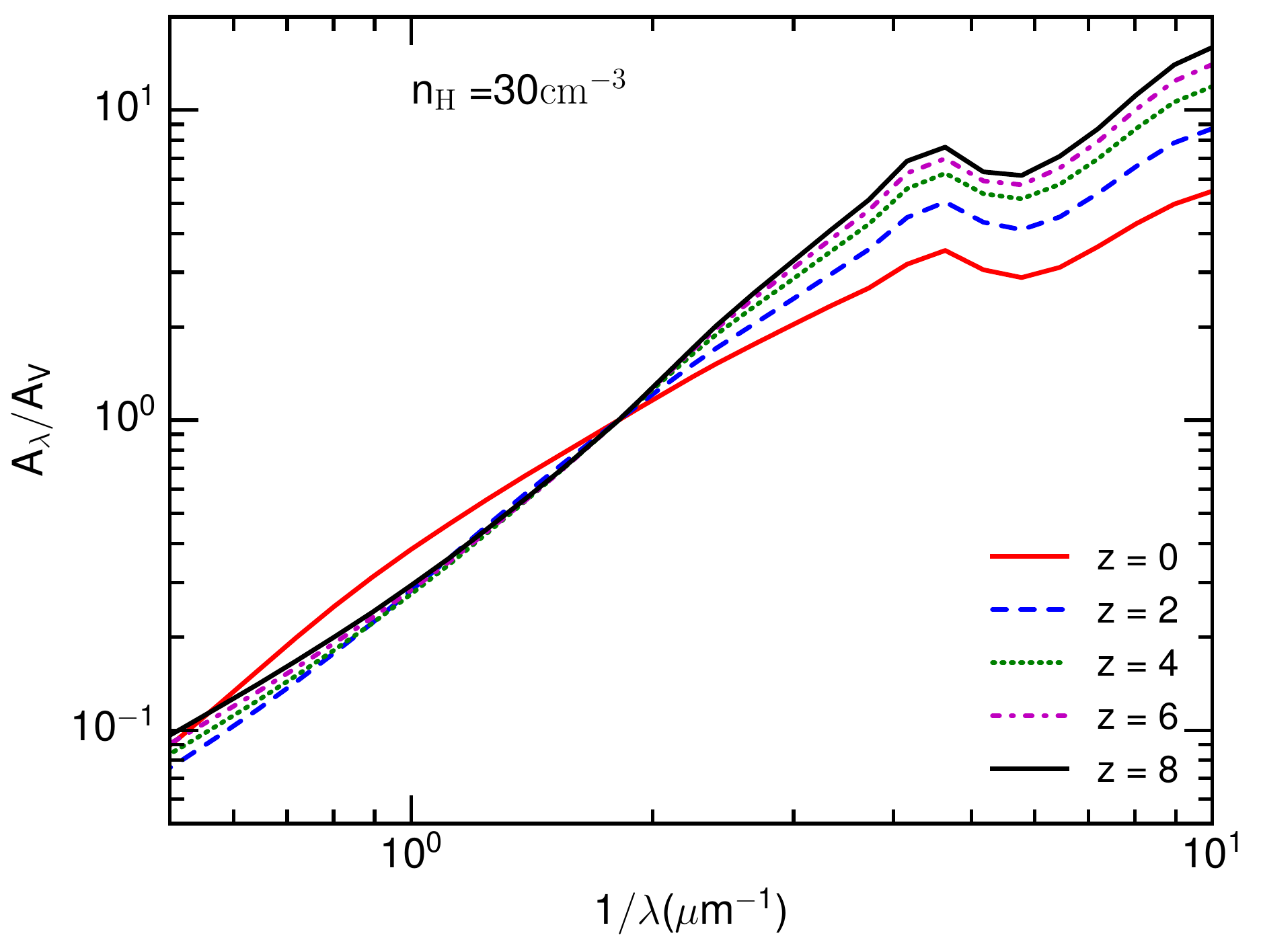}
\includegraphics[width=0.5\textwidth]{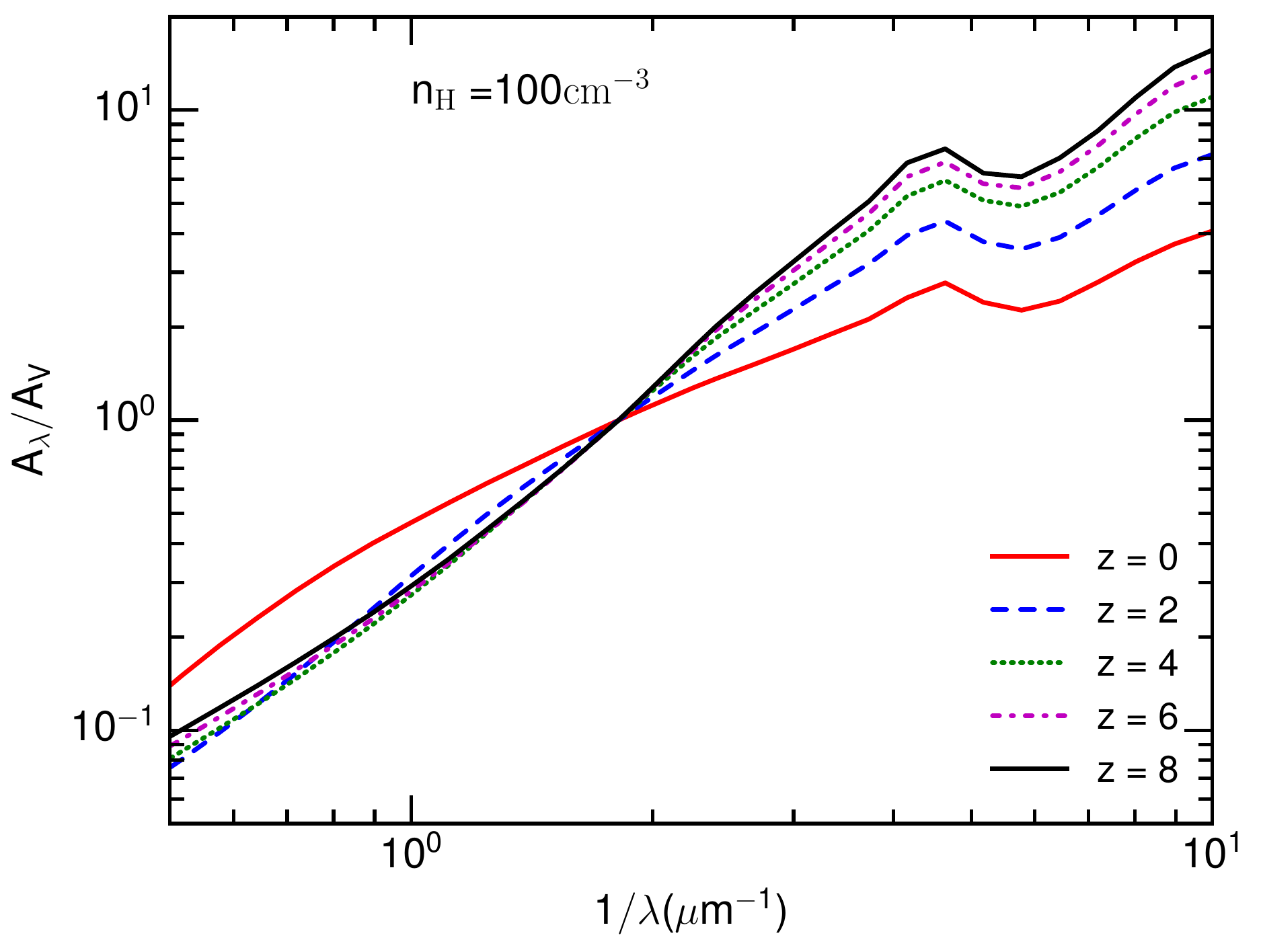}
\caption{Normalized extinction curves ($A_{\lambda}/A_{V}$), for different redshift ($z$) and gas density ($n_{\H}$), assuming the typical tensile strength $S_{\max}=10^{5}\erg\cm^{-3}$ for composite grains. UV extinction increases and optical-NIR extinction decreases with redshift due to disruption of large grains into smaller ones, resulting in steeper extinction curves.}
\label{fig:Aext_S1e5}
\end{figure*}

Figure \ref{fig:Aext_S1e5} shows the normalized extinction curves ($A_{\lambda}/A_{V}$) at different redshift for different gas density $n_{\H}=1-100\cm^{-3}$ and $T_{\gas}=100\K$, assuming that grains have composite structure with $S_{\max}=10^{5}\erg\cm^{-3}$. In general, the UV extinction increases while the optical-NIR extinction decreases with redshift, resulting in steeper extinction curves. This arises from the effect of RATD that breaks large grains of $a>a_{\rm disr}$ into smaller ones. Indeed, since larger grains have higher contribution to extinction at optical-NIR wavelength and small grains contribute more to UV extinction, the conversion of large to small grains decreases optical-NIR extinction and increases UV extinction. 
One can see that the magnitude of the increase in UV extinction with redshift appears to be stronger for higher density. This can be seen from the variation of disruption size with $z$ in Figure \ref{fig:adisr} where a larger variation in $a_{\rm disr}$ is seen for larger $n_{\H}$.

\begin{figure*}
\includegraphics[width=0.5\textwidth]{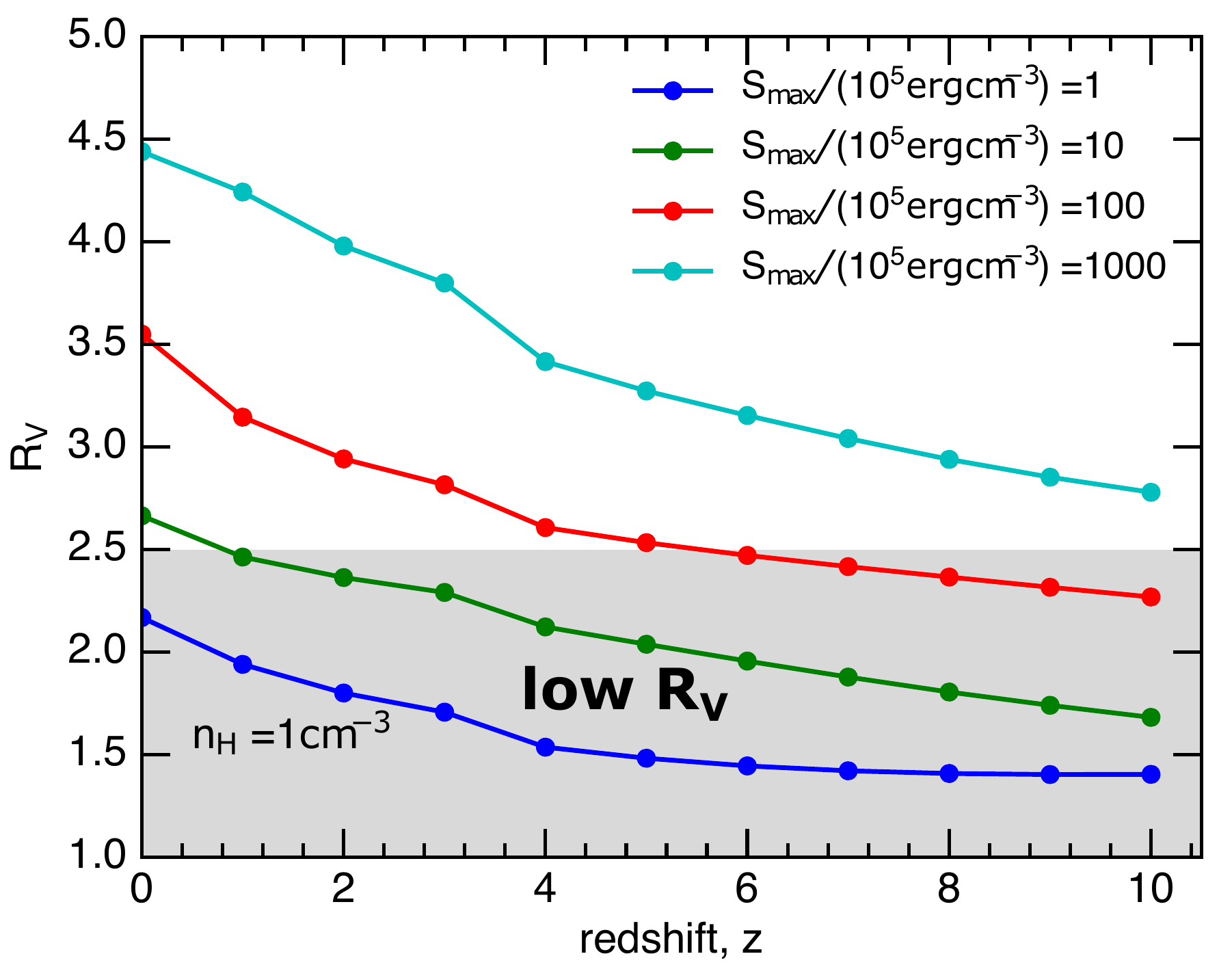}
\includegraphics[width=0.5\textwidth]{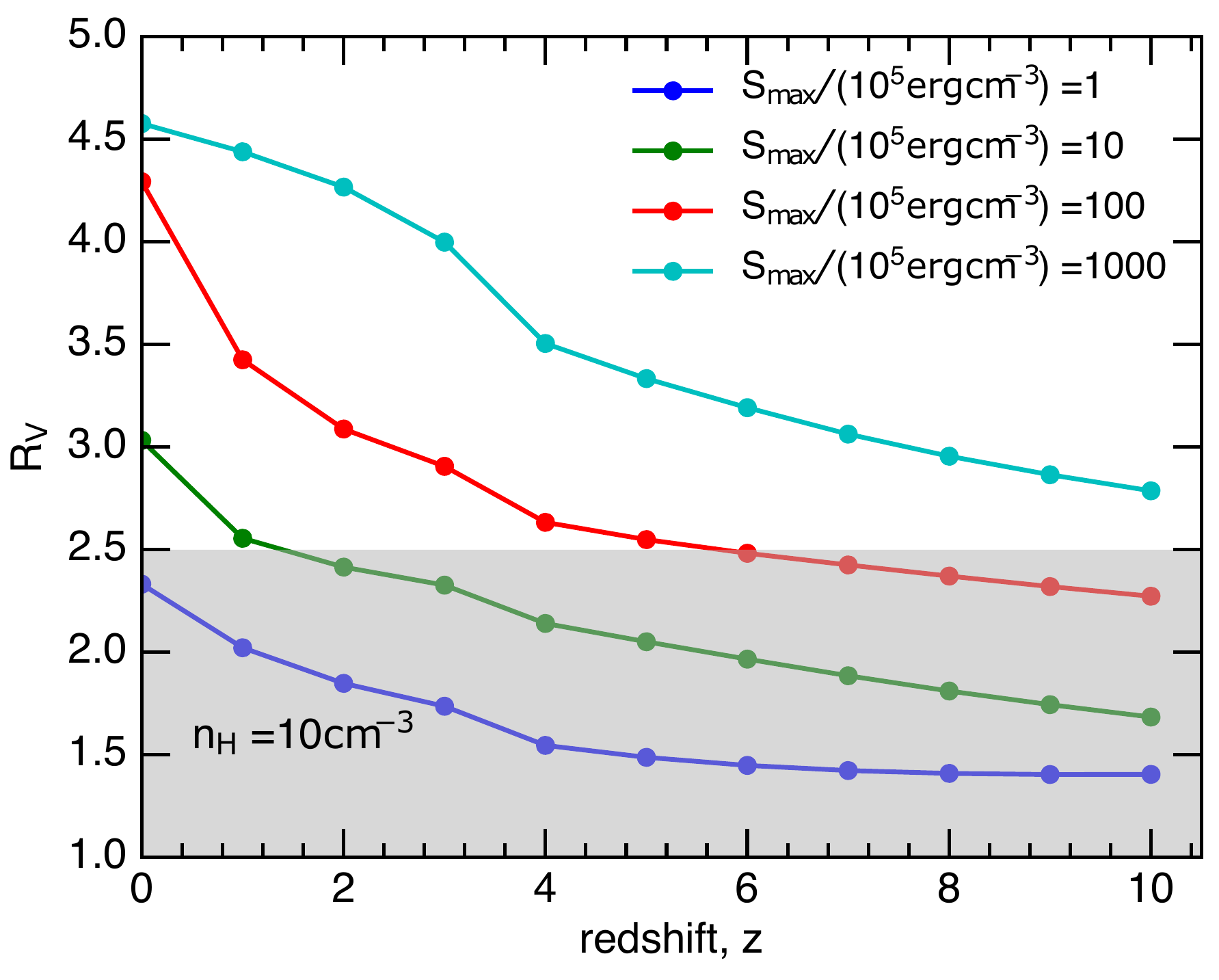}
\includegraphics[width=0.5\textwidth]{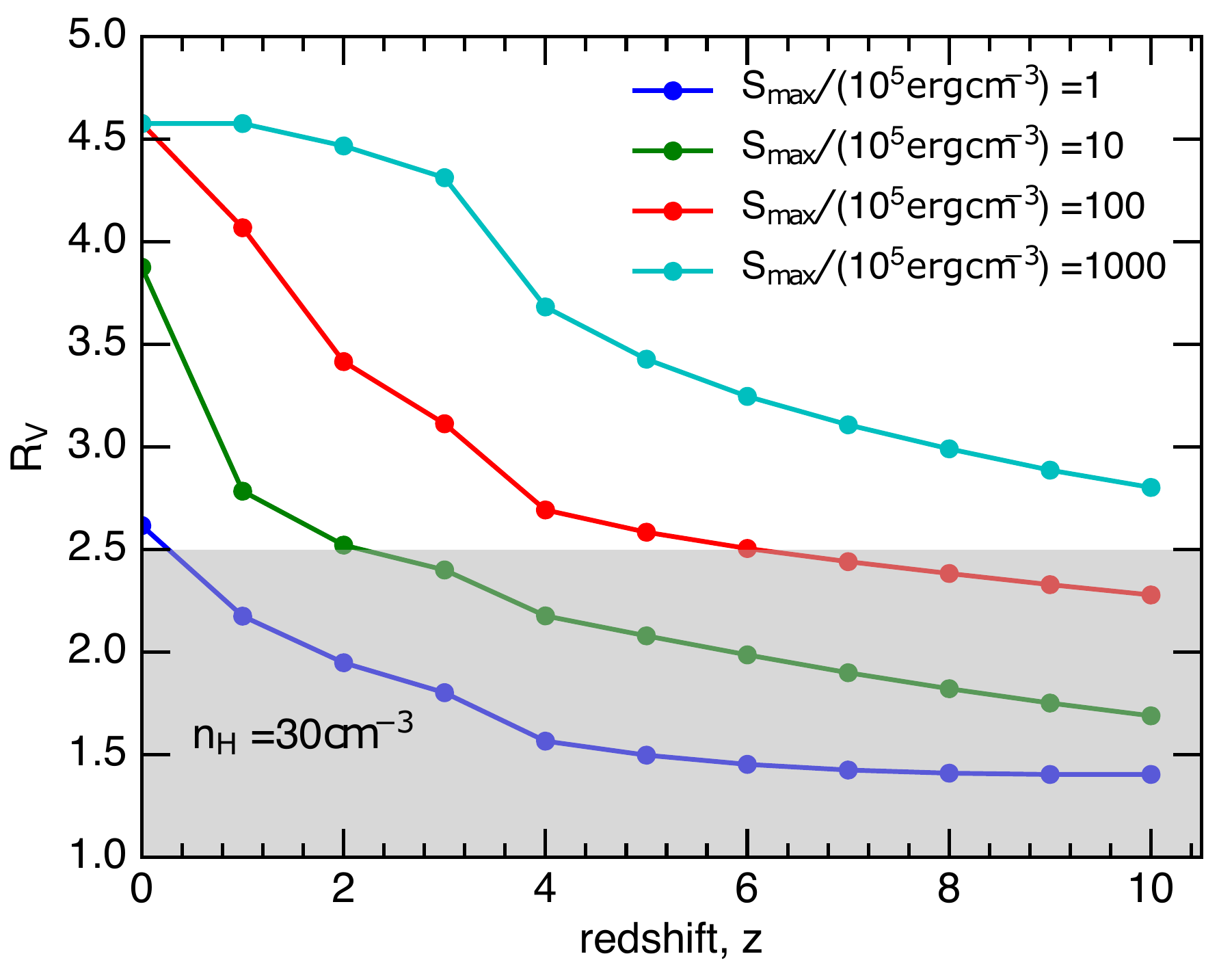}
\includegraphics[width=0.5\textwidth]{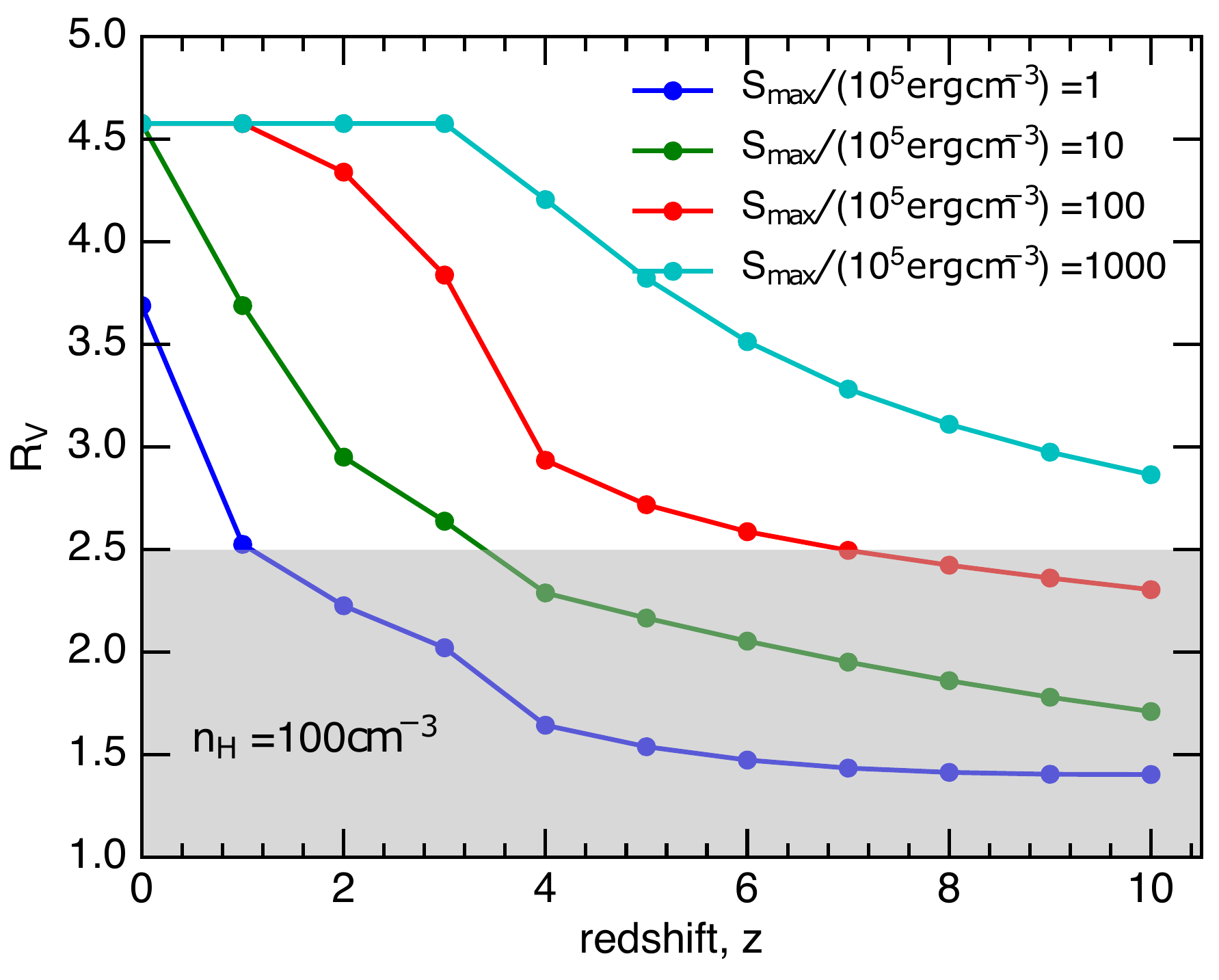}
\caption{Decrease of $R_{V}$ with redshift for the different tensile strengths and local gas density. Shaded area represents the low value of $R_{V}<2.5$. Composite grains of $S_{\max}=10^{5}\erg\cm^{-3}$ induces low $R_{V}<2.5$ in all redshift and falls to $R_{V}\sim 1.5$ for $z>4$.}
\label{fig:RV_disr}
\end{figure*}


Using the obtained extinction curves, we calculate the ratio of total-to-selective extinction, $R_{V}=A_{V}/(A_{B}-A_{V})$.
Figure \ref{fig:RV_disr} shows the decreases of $R_{V}$ with $z$ for different $S_{\max}$ and gas density. The shaded regions highlight low values of $R_{V}<2.5$.

For a given density, $R_{V}$ rapidly decreases with redshift. At a given redshift, $R_{V}$ decreases with decreasing $n_{\H}$. For the typical strength of composite grains ($S_{\max}=10^{5}\erg\cm^{-3}$), one obtains $R_{V}<2.5$ for the considered gas densities, decreasing from $R_{V}\sim 2-3.5$ at $z=0$ to $R_{V}\sim 1.5$ for $z>4$ (see blue lines). For grains of strong material with $S_{\max}=10^{8}\erg\cm^{-3}$ (e.g., compact structure), one also see the decrease of $R_{V}$ with redshift, from $R_{V}\sim 5$ at $z=0$ to $R_{V}< 3.1$ at $z>6$. For the diffuse medium with $n_{\H}<30\cm^{-3}$, composite grains all have low $R_{V}<2$ at high-z. We find that the decrease of $R_{V}$ occurs rapidly for $z=0-5$. Above $z=5$, the decrease of $R_{V}$ is slower because radiation intensity become sufficiently large that IR damping becomes dominant, and the disruption size slowly decreases with $U$ (see Eq. \ref{eq:adisr_nogas1}). 

\begin{figure*}
\includegraphics[width=0.5\textwidth]{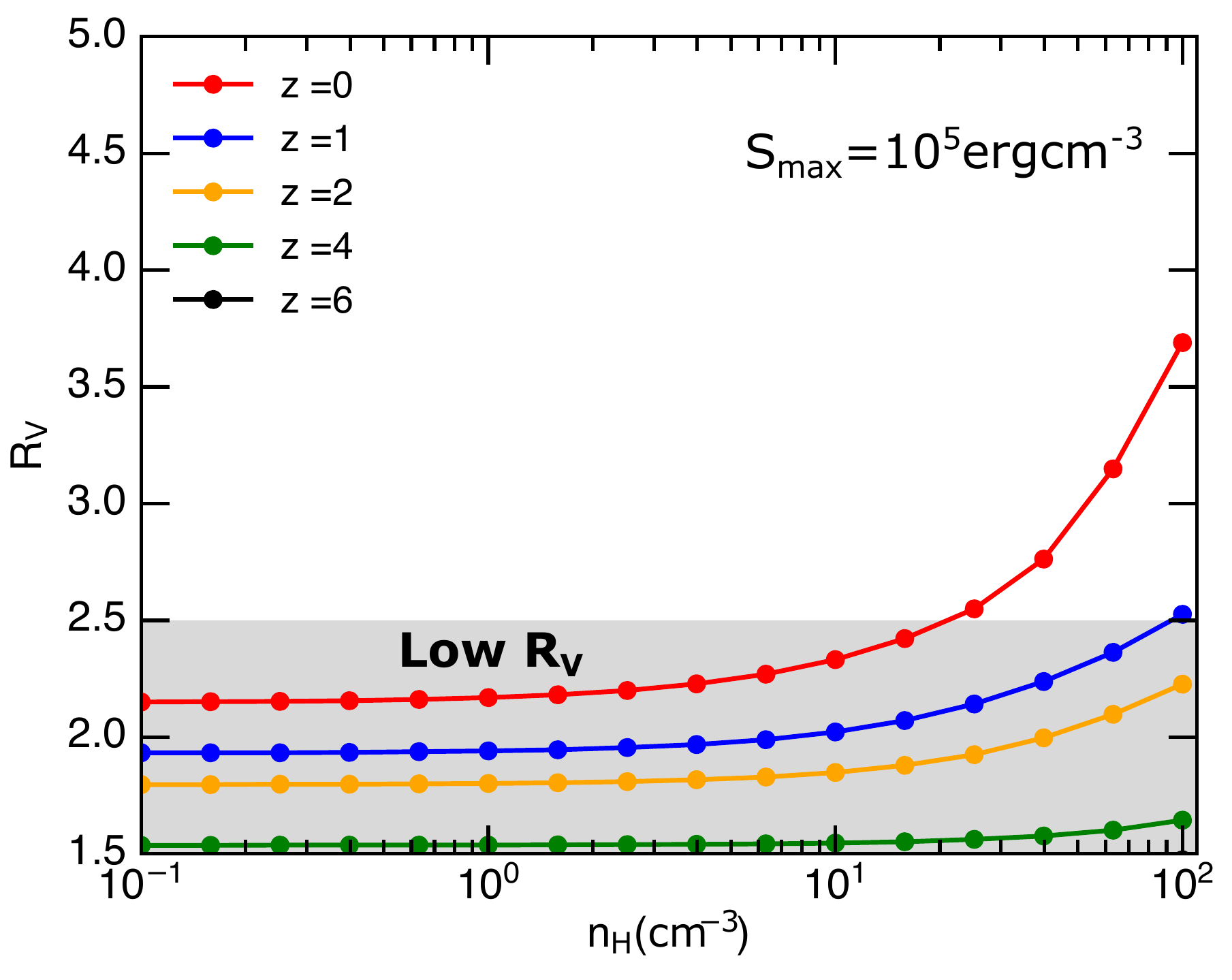}
\includegraphics[width=0.5\textwidth]{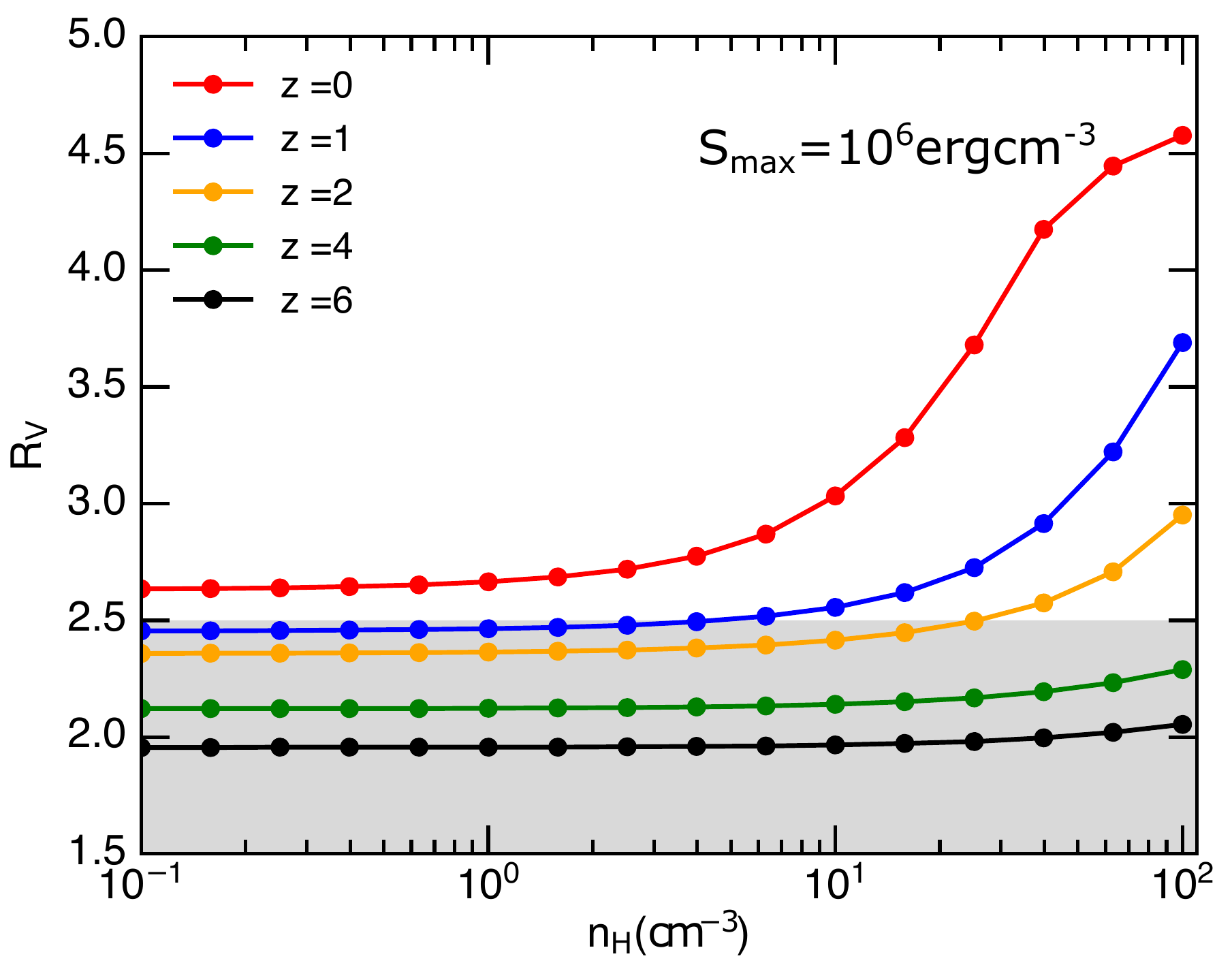}
\includegraphics[width=0.5\textwidth]{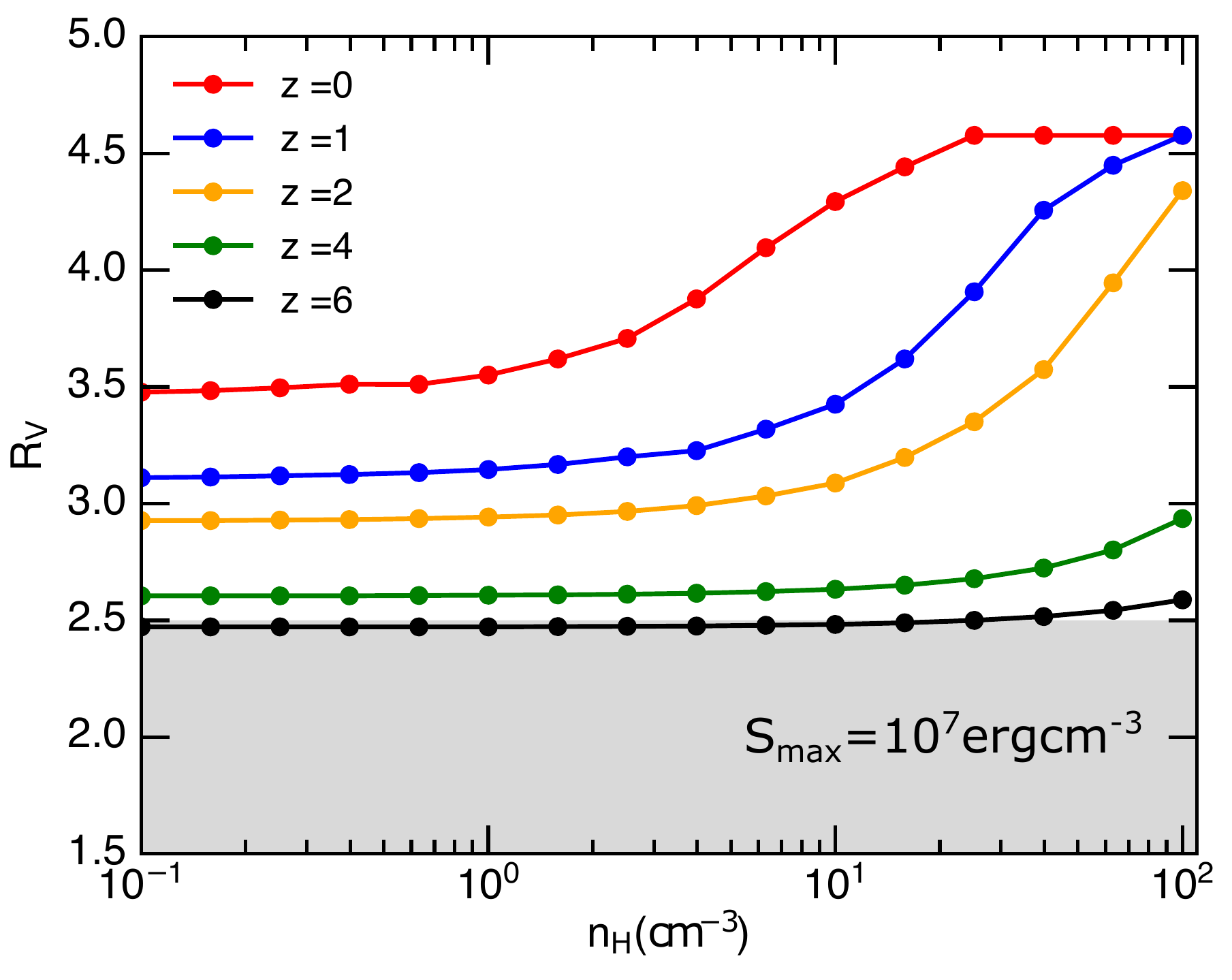}
\includegraphics[width=0.5\textwidth]{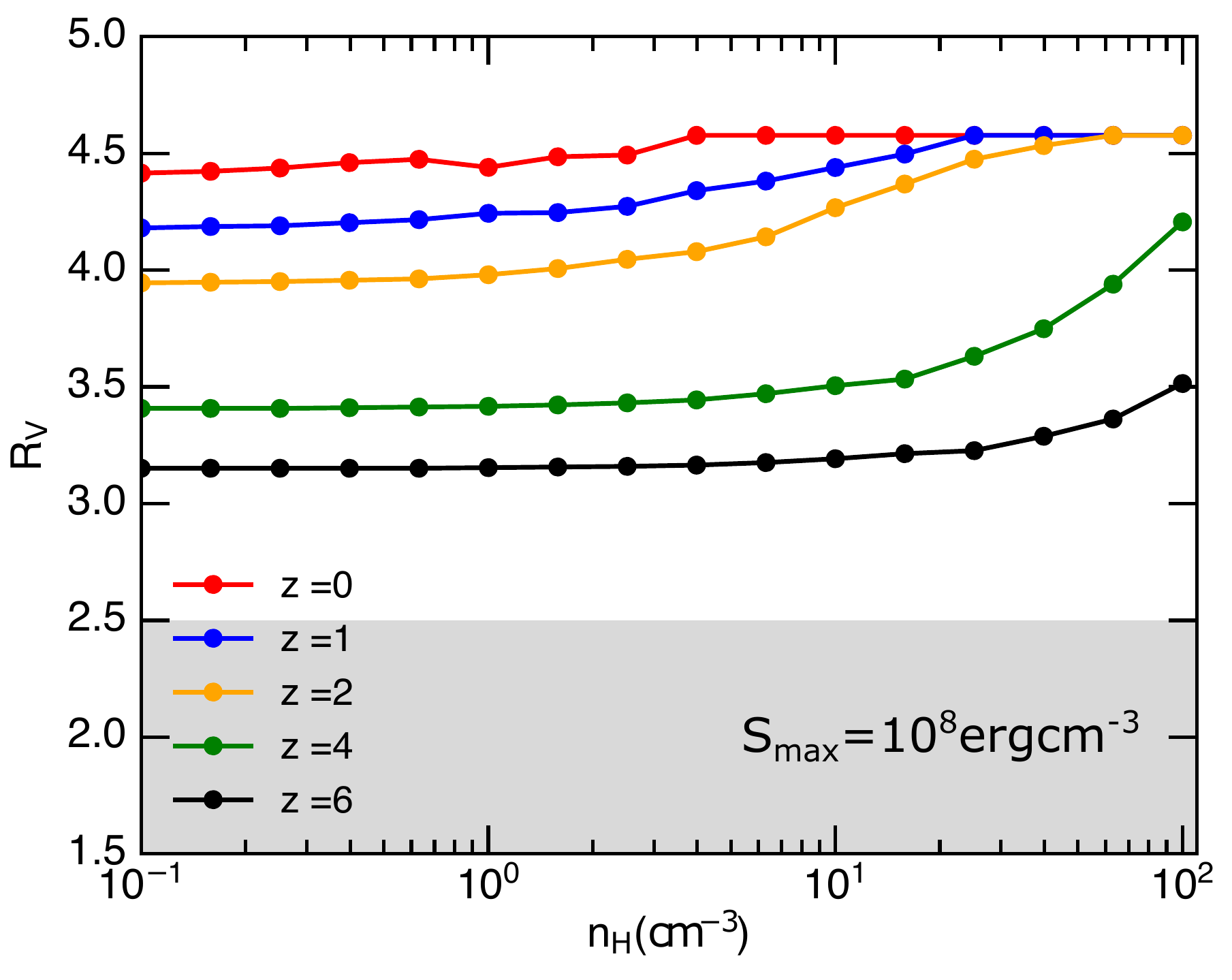}
\caption{Variation $R_{V}$ with the local gas density for different redshifts, assuming $S_{\max}=10^{5}-10^{8}\erg\cm^{-3}$. Shaded area marks the region of low $R_{V}<2.5$. The variation is most sensitive for $z\sim 0-1$ and becomes insensitive at high $z$ when the mean radiation intensity becomes sufficiently large.}
\label{fig:RV_nH}
\end{figure*}

In Figure \ref{fig:RV_nH} we show the variation of $R_{V}$ with $n_{\H}$ for different redshift and tensile strength. The value of $R_{V}$ increases with $n_{\H}$ for $z<4$, but it becomes saturate at higher $z$ due to the high radiation intensity that makes disruption independent of $n_{\H}$. For $z\sim 0-2$, the value $R_{V}$ is almost constant for $n_{\H}<1\cm^{-3}$ and starts to increase rapidly for $n_{\H}\gtrsim 1\cm^{-3}$. At larger $z$, the variation of $R_{V}$ is less pronounced in the considered range of gas density due to dominance of IR damping over gas damping.

\subsection{Effect of varying tensile strength with grain structure}
\begin{figure*}
\includegraphics[width=0.5\textwidth]{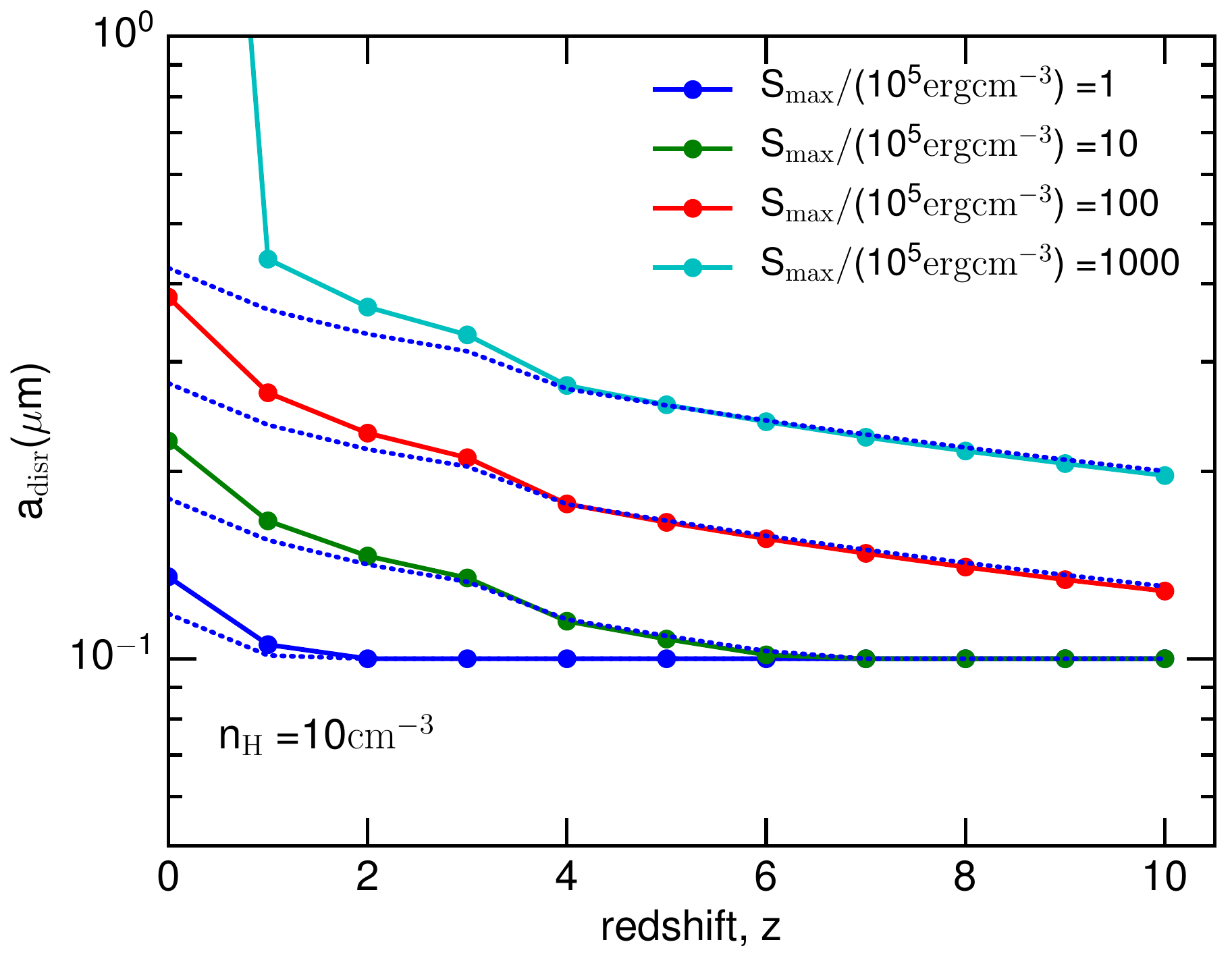}
\includegraphics[width=0.5\textwidth]{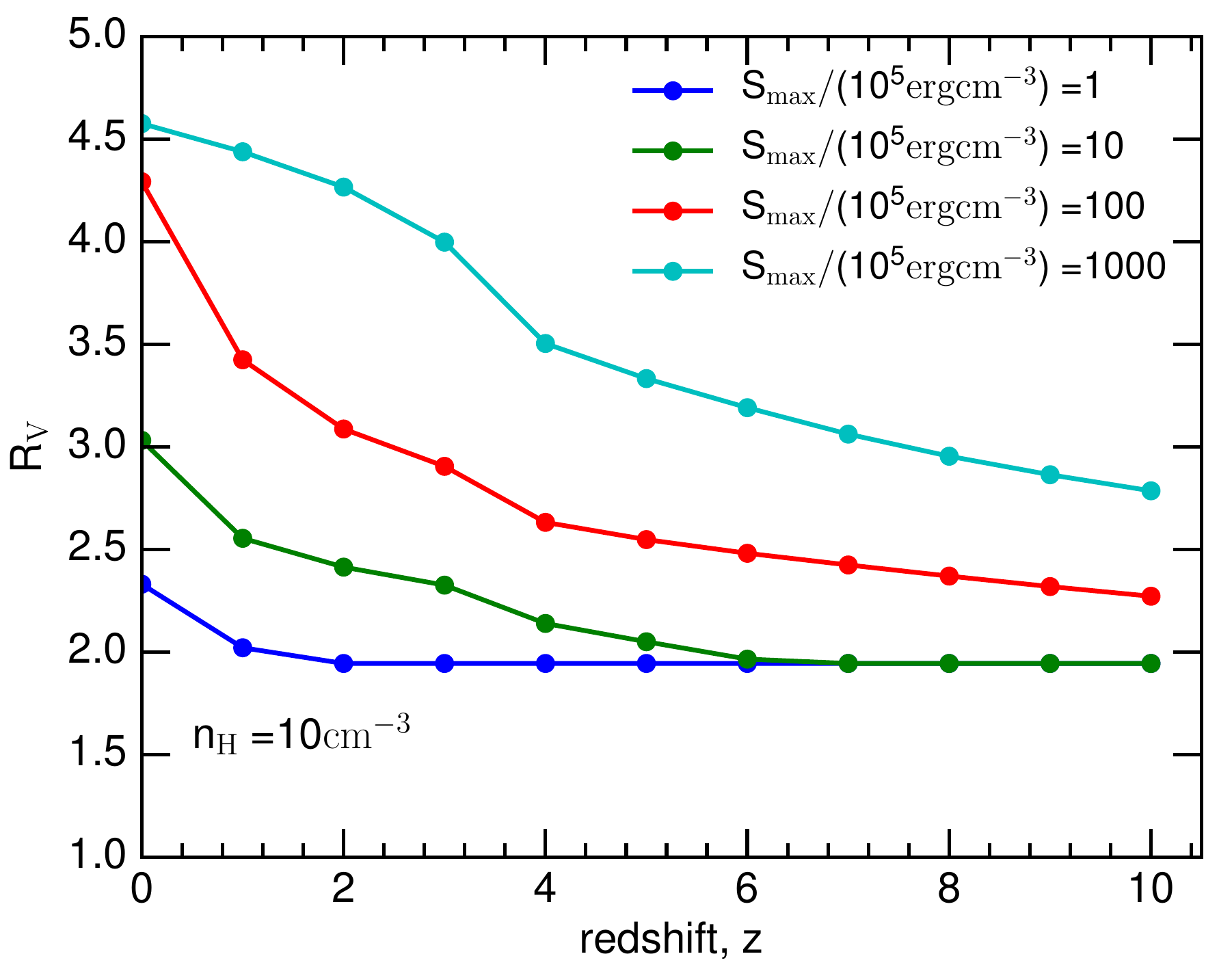}
\caption{Variation of disruption size (left) and $R_{V}$ (right panel) with redshift when grains smaller than $a=0.1\mum$ are assumed to be compact of $S_{\max}=10^{9}\erg\cm^{-3}$. The disruption size reaches its minimum of $0.1\mum$ (left panel). The value of $R_{V}$ decreases rapidly to its saturated value at $R_{V}=1.945$ for two cases of lowest $S_{\max}$. The gas density $n_{\H}=10\cm^{-3}$ is assumed for the ISM.}
\label{fig:RV_Svary}
\end{figure*}

We now assume that grains smaller than $a=a_{\rm core}=0.1\mum$ are compact and have a high tensile strength of $S_{\max}=10^{9}\erg\cm^{-3}$. Therefore, the disruption size cannot go below $a_{\rm disr}=a_{\rm core}$ as shown in the left panel of Figure \ref{fig:RV_Svary}. We then run calculations of extinction curves and obtain $R_{V}$ as shown in the right panel of Figure \ref{fig:RV_Svary}. The value of $R_{V}$ decreases rapidly to its saturated value at $R_{V}=1.945$ for two cases of low strength of $S_{\max}=10^{5}$ and $10^{6}\erg\cm^{-3}$. Grains with larger $S_{\max}$ cannot be disrupted down to the core radius, so $R_{V}$ does not change.

\section{Discussion}\label{sec:disc}
\subsection{Smaller grains at higher redshifts}
We have applied the RATD mechanism to study the variation of the grain size distribution across cosmic time. Since the grain internal structures that determine the tensile strength are uncertain, we consider a range of tensile strength between $10^{5}-10^{8}\erg\cm^{-3}$, which are implied by composite or core-mantle structures. We note that such structures are expected for interstellar dust due to various destruction and coagulation processes between the diffuse ISM and dense clouds (\citealt{1990ARA&A..28...37M}; \citealt{Draine:2020ua}).

Using the fact that the mean intensity of interstellar radiation increases with redshift (plausibly due to a higher star formation efficiency), we find that the disruption size that determines the maximum size of grains decreases rapidly with $z$. Depending on the internal structure of grains, the maximum size can be larger for more compact grains of larger $S_{\max}$. If grains have a typical composite structure of $S_{\max}=10^{5}\erg\cm^{-3}$, their maximum size is largest of $a\sim 0.2\mum$ at $z=0$ and rapidly decreases to $a\le 0.1\mum$ at higher $z$. Therefore, dust at higher $z$ is consequently dominated by small grains. Nevertheless, the presence of larger grains at high-z is not ruled out if they have a compact structure.

The disruption size also varies with the local gas density $n_{\H}$, but it becomes independent of $n_{\H}$ for $z>4$ when the radiation intensity is sufficiently strong for rotational damping by IR emission becomes dominant over the gas damping at $n_{\H}\lesssim 30\cm^{-3}$. We note that grain disruption is inefficient for dense molecular clouds without embedded stars where grain growth is driven by coagulation \citep{2013MNRAS.434L..70H}. If molecular clouds contain embedded sources, rotational disruption is still efficient in proximity of the sources, as found in \cite{Hoang:2020vg}.

\subsection{Steeper extinction curves, smaller $R_{V}$, and implications for observations}
Due to RATD, large grains are converted into smaller ones. As a result,
the extinction curve becomes steeper with increasing redshift (see Figure \ref{fig:Aext_S1e5}). 

The ratio of total-to-selective extinction, $R_{V}$, is found to decrease with redshift. For composite grains, one has $R_{V}<2.5$ in the diffuse regions with $n_{\H}\le 100\cm^{-3}$ (see Figure \ref{fig:RV_disr}). For grains of higher tensile strength, $R_{V}$ could be larger than the standard $R_{V}=3.1$, but its decreasing trend is observed. At $z=0$, composite grains are not disrupted for dense regions of $n_{\H}> 50\cm^{-3}$ and $R_{V}$ increases to larger than $3.1$. However, at high-z, disruption is still efficient at such high density and $R_{V}$ only exceeds $3.1$ if grains are compact (see Figure \ref{fig:RV_nH}).

In starburst galaxies with star formation activities and supernova explosions, observations usually show peculiar extinction curves with a steep far-UV rise, SMC-like curve \citep{1997ApJ...487..625G}. The SMC-like extinction curves are also observed toward the host galaxies of gamma-ray-bursts (\citealt{2012A&A...537A..15S}; \citealt{Heintz:2019ey}), quasars \citep{2004AJ....128.1112H}, and high-z star-forming galaxies \citep{2018ApJ...853...56R}. Our theoretical modeling of extinction induced by RATD implies steep extinction curves with $R_{V}<2.5$ for $z>6$ if grains do not have compact structure with tensile strength of $S_{\max}\gtrsim 10^{8}\erg\cm^{-3}$ (see Figure \ref{fig:RV_disr}), which successfully reproduce the SMC-like extinction curves observed toward high-z galaxies using quasars or gamma ray bursts.


\subsection{Origins of anomalous dust extinction toward SNe Ia}
Extinction curves toward SNe Ia are known to be anomalous, with unusually low value of $R_{V}\sim 1-2.5$ with the mean value of $\langle R_{V}\rangle \approx 1.7$ (\citealt{2014ApJ...789...32B}; \citealt{2016ApJ...819..152C}) and $\langle R_{V}\rangle\approx 2.71$ \citep{2016ApJ...819..152C}. The exact origin of such low $R_{V}$ is unknown. \cite{2008ApJ...686L.103G} suggested a multiple scattering model by circumstellar dust as a cause of low $R_{V}$. Yet, the non-detection of NIR emission from SN 2014J by {\it Spitzer} (\citealt{2017MNRAS.466.3442J}) as expected from hot circumstellar dust casts doubt on this scenario. Rotational disruption of grains by RATs proposed by \cite{Hoang:2019da} could reproduce such low values based on disruption induced by SNe flash (\citealt{2020ApJ...888...93G}). In this scenario, there must exist a dust cloud within 4 pc from the source. The unique prediction of disruption by SNe flash is the variation of extinction and polarization of SNe light with time. 

Our results here show that if grains have a composite structure and are located in the diffuse environment ($n_{\H}\le 30\cm^{-3}$), $R_{V}$ is small, $\sim 1.9-2.5$ for $z\sim 0-1$ (see Figure \ref{fig:RV_disr}), which adequately explain the estimated low $R_{V}$ values of SNe Ia. In particular, some SNe, including SNe 2006X, 2008fp, 2014J, exhibit extreme values of $R_{V}<1.5$ (see \citealt{2017ApJ...836...13H}). The adopted radiation intensity for main sequence galaxies cannot produce these extreme values. If the mean intensity of these galaxies is enhanced by starburst for which the grain temperature can reach $T_{d}\sim 60\K$ ($U\sim 10^{3}$) (\citealt{Zavala:2018cn}), corresponding to $z\sim 9$ if Equation (\ref{eq:Td_z}) is used. For this radiation intensity, $R_{V}$ can be reproduced by RATD if grains have composite structures (see Figure \ref{fig:RV_disr}). Note that, due to its interstellar nature, extinction and polarization curves implied by RATD do not vary with the observational time (cf. disruption by SNe light, \citealt{2020ApJ...888...93G}). 

\subsection{Space and time variation of $R_{V}$ and Implications for SNe cosmology}
The well-known crisis in cosmology is the tension in measurements of the Hubble constant using SNe Ia and Cosmic Microwave Background (CMB) radiation. SNe Ia measurements report $H_{0}=74.03\pm 1.42 {\rm km} \s^{-1} {\rm Mpc}^{-1}$ (\citealt{2019ApJ...876...85R}), whereas CMB measurements by \cite{2020A&A...641A...6P} report $H_{0}=67.4\pm 0.5 {\rm km} \s^{-1} {\rm Mpc}^{-1}$. Moreover, \cite{Freedman:2019fn} report $H_{0}=69.8\pm 0.8 {\rm km} \s^{-1} {\rm Mpc}^{-1}$ using on the Tip of the Red Giant Branch.

Dust extinction is critical to the precise measurements of H$_0$ using SNe Ia standardized candles. \cite{Brout:2020ww} suggested that the scatter in $H_{0}$ could be completely reproduced by allowing the variation of $R_{V}$. Recent analysis in \cite{GonzalezGaitan:2020vd} also confirms the importance host galaxies dust. Since most of SNe Ia are expected to explode in the low density diffuse medium, it is unclear what causes the variation of $R_{V}$. 

Our modeling results in Figure \ref{fig:RV_disr} show that, due to rotational disruption by RATs, $R_{V}$ varies with the physical parameters of the local environment, including the gas density and the radiation field. It also changes with the grain structures. It is known that the ISM is turbulent (\citealt{Armstrong:1995p3264}), producing fluctuations in the gas density. Moreover, the local intensity depends on the distribution of stars. Therefore, the value of $R_{V}$ experiences strong fluctuations along the different lines of sight in the host galaxy. Such fluctuations in $R_{V}$ inevitably induce the scatter in the inferred measurements of the Hubble constant.

\subsection{Implications for high-z astrophysics}
ALMA is revolutionizing our research on dust and gas in early universe up to redshift $z\sim 10$ (see \citealt{Bouwens:2020tf}). The relationship between the infrared (IRX) and UV slope ($\beta$), namely IRX-$\beta$, of the spectral energy density (SED) is a key parameter to estimate the star-formation rate (SFR) in galaxies (e.g., \citealt{2016ApJ...833...72B}). Therefore, an accurate extinction curve is critically important for reliable estimates of SFR. In light of our study, the extinction curves at high-z are steeper than the standard Milky Way, and the steepness increases with redshift. Therefore, it poses a challenge to an accurate determination of star-formation activity in early universe.


\section{Summary}\label{sec:summ}
We study the variation of dust properties with redshift resulting from rotational disruption by RATs (RATD mechanism) induced by interstellar radiation field (ISRF). The main results are summarized as follows:

\begin{itemize}

\item{} The efficiency of RATD increases with redshift due to the increase in the mean radiation intensity. The maximum size of the grain size distribution thus decreases with increasing redshift but increases with the gas density. For $z>4$, the disruption size of dust in the diffuse medium becomes independent of the density.

\item{} Rotational disruption converts large grains into smaller, thus, grains become smaller at higher redshifts. Resulting extinction curves become steeper, and the ratio of total-to-selective extinction, $R_{V}$, decreases rapidly with redshift.

\item If grains have composite structures of tensile strength $S_{\max}\lesssim 10^{6}\erg\cm^{-3}$, $R_{V}$ is small, between $1.5-2.5$, much smaller than the standard value of $R_{V}=3.1$ in the Galaxy. This can reproduce the popular SMC-like extinction curves observed toward high-z galaxies.

\item Unusually low values of $R_{V}\sim 1.5-2.5$ observed toward SNe Ia of $z<1$ could be reproduced by RATD induced by ISRF if grains have composite structures, but the extreme values of $R_{V}<1.5$ observed for several SNe Ia
require an enhanced radiation field. Alternatively, it can be reproduced by RATD if there exist some nearby clouds within several parsecs could.

\item{} The fluctuations of $R_{V}$ due to variation of the local gas density by interstellar turbulence, radiation intensity, and redshift inevitably affect the accurate measurements of the Hubble constant $H_{0}$. This might help to resolve the tension between local SNe measurements and early measurements using CMB. 

\item{} The variation of dust properties also affects the star-formation rate measured toward high-z galaxies. Thus, one should account for the variation of extinction curves with $z$ to achieve accurate measurements.

\end{itemize}

\acknowledgments
We thank the anonymous referee for a helpful report, and A. Goobar and M. Bulla for fruitful conversation on dust properties toward SNe Ia. T.H. acknowledges the support by the National Research Foundation of Korea (NRF) grants funded by the Korea government (MSIT) through the Mid-career Research Program (2019R1A2C1087045).


\bibliography{ms.bbl}

\end{document}